\documentclass[a4paper,10pt,twoside]{article}

\usepackage{appendix}
\usepackage{hyperref}
\usepackage{psfrag}
\usepackage{chemarr}
\usepackage{appendix}
\usepackage{subfigure}
\usepackage{amssymb}
\usepackage{multirow}
\usepackage{amsmath}
\usepackage{xr}
\usepackage{psfrag}
\usepackage{array}
\usepackage{placeins}

\newcommand{\ecoli}{{\em E.coli}}
\newcommand{\sm}{{\em Appendix}}

\newcommand{\onel}{($A$)}
\newcommand{\twol}{($B$)}
\newcommand{\threel}{($C$)}
\newcommand{\fourl}{($D$)}
\newcommand{\fivel}{($E$)}
\newcommand{\sixl}{($F$)}
\newcommand{\sevenl}{($G$)}
\newcommand{\eightl}{($H$)}

\usepackage{graphicx}

\usepackage[numbers,round,sort&compress]{natbib}

\usepackage{hyperref}

\usepackage[margin=2.5cm]{geometry}

\title{Physical constraints determine the logic of bacterial promoter architectures}
\author{Daphne Ezer$^{1,2,\ast}$, Nicolae Radu Zabet$^{1,2,\ast,\ddagger}$ and Boris Adryan$^{1,2,\dagger}$    \\
\mbox{}\\
\footnotesize $^1$ Cambridge Systems Biology Centre, University of Cambridge, Tennis Court Road, Cambridge CB2 1QR, UK;\\
\footnotesize $^2$ Department of Genetics, University of Cambridge, Downing Street, Cambridge CB2 3EH, UK\\
\footnotesize $^\ast$ D.E. and N.R.Z. contributed equally to this work.\\
\footnotesize $^\dagger$Corresponding author: adryan@sysbiol.cam.ac.uk, \footnotesize $^\ddagger$Correspondence can also be addressed to: n.r.zabet@gen.cam.ac.uk}
\date{}

\begin{document}

\maketitle

\begin{abstract}
Site-specific transcription factors (TFs) bind to their target sites on the DNA, where they regulate the rate at which genes are transcribed. Bacterial TFs undergo facilitated diffusion (a combination of 3D diffusion around and 1D random walk on the DNA) when searching for their target sites. Using computer simulations of this search process, we show that the organisation of the binding sites, in conjunction with TF copy number and binding site affinity, plays an important role in determining not only the steady state of promoter occupancy, but also the order at which TFs bind. These effects can be captured by facilitated diffusion-based models, but not by standard thermodynamics. We show that the spacing of binding sites encodes complex logic, which can be derived from combinations of three basic building blocks: switches, barriers and clusters, whose response alone and in higher orders of organisation we characterise in detail. Effective promoter organizations are commonly found in the \emph{E. coli} genome and are highly conserved between strains. This will allow studies of gene regulation at a previously unprecedented level of detail, where our framework can create testable hypothesis of promoter logic.
\end{abstract}

\section*{Introduction}
Bacterial promoters are often very complex, containing many densely spaced and potentially overlapping transcription factor (TF) binding sites \cite{hermsen_2006}. The rate of gene expression depends on the promoter configuration (the specific combination of TFs that are bound to the promoter), and specific rules (``logic'') may simply depend on the presence of two physically interacting TFs. Here, we propose that the dynamics of TF binding can influence promoter occupancy over time and therefore provide a time-dependent trigger that determines how TFs can depend on binding site spacing to influence gene expression.
 
One common method of identifying where TFs bind is to search a DNA sequence for TF binding site motifs, as specified by Position Weight Matrices (PWMs) \cite{stormo_2000}.  Frequently, PWMs are used alongside statistical thermodynamic-based methods in order to incorporate additional properties influencing TF binding, such as TF concentration and spatial hindrance between TFs \cite{ackers_1982,djordjevic_2003,bintu_2005_model,foat_2006,roider_2007,chu_2009,zhao_2009,sadka_2009,wasson_2009,hoffman_2010,kaplan_2011,simicevic_2013}. 
 
These thermodynamic ensemble models assume that the probability of a configuration occurring is directly correlated with the thermodynamic stability of that configuration, which is primarily influenced by the binding site affinities and protein abundances.  However, some thermodynamically stable configurations may take a long time to form, thereby decreasing the likelihood that those configurations occur within the time frame of a cell cycle. In order to model promoter configuration without requiring strong assumptions about the presence of thermodynamic equilibrium, the kinetics underlying TF binding must be taken into account.

Both \emph{in vitro} and \emph{in vivo} studies have shown  that TFs find their sites by facilitated diffusion \cite{riggs_1970a,kabata_1993,blainey_2006,elf_2007,hammar_2012}; note that reference \cite{hammar_2012} provided strong evidence that TFs use facilitated diffusion as a translocation mechanism \emph{in vivo}. This mechanism assumes that proteins do not home in on their target sites by 3D diffusion alone, but also take a random walk linearly along the DNA, in effect reducing the dimensionality of the search to find their binding sites more efficiently \cite{berg_1981,halford_2004,mirny_2009,zabet_2012_review,kolomeisky_2011}.

There have been many attempts to mathematically analyse the facilitated diffusion mechanism using analytical solutions \cite{berg_1981,halford_2004,coppey_2004,slutsky_2004b,sokolov_2005,klenin_2006,hu_2006,benichou_2008,li_2009,lomholt_2009,loverdo_2009,meroz_2009,vukojevic_2010,benichou_2011,zhou_2011}. However, these mathematical approximations frequently assume an uniform affinity landscape and do not capture the stochastic behaviour of the system.  We have previously established a stochastic simulation framework called GRiP (Gene Regulation in Prokaryotes) that can incorporate real affinity landscapes and therefore provides more accurate predictions of TF binding kinetics  \cite{zabet_2012_grip,zabet_2012_model,zabet_2012_review}.  \emph{In vivo} single-molecule microscopy experiments have been used to measure various physical parameters in the facilitated diffusion process of the \emph{E. coli} TF \emph{lacI}, including the average length of time \emph{lacI} is bound to the DNA during its random walk, the average distance \emph{lacI} traverses during its random walk, and the proportion of time \emph{lacI} is undergoing 1D diffusion versus 3D diffusion \cite{elf_2007,hammar_2012}. We derive all of the kinetic parameters in our simulations from these measured experimental values \cite{zabet_2012_model}; see Table \ref{tab:paramsDescription} in the \sm.  
 
Transcriptional logic refers to the idea that the output -- the expression level of a gene -- depends on the specific combination of multiple inputs, the concentrations of TFs that regulate that gene. Typically, one considers the system to be in steady state, with the binding of the TFs to the promoter to be in quasi-equilibrium \citep[e.g.][]{bintu_2005_model,hermsen_2006,sadka_2009}. Here we extend this notion by proposing that the response to multiple inputs can also depend on the kinetics of TF binding, e.g. on the order by which the TFs bind to the promoter. In this context, we suggest that the spatial organisation of the promoter encodes the logic of how TF concentration influences the promoter occupancy dynamics in biologically relevant time scales. Based on the facilitated diffusion model, we identified three basic functional units of diffusion-based transcriptional logic: $(i)$ the switch (two overlapping TF binding sites), $(ii)$ the barrier (two closely spaced, but non-overlapping TF binding sites) and $(iii)$ the cluster (two closely spaced or overlapping binding sites for the same TF). Furthermore, we utilize the behaviours of these promoter building blocks to develop a semi-analytical model of the facilitated diffusion mechanism, which is significantly less resource-intensive than fully stochastic simulations and thus allows for genome-wide investigation.  We then systematically describe the theoretical behaviour of these building blocks across possible concentrations and binding affinities and demonstrate that combining these building blocks can result in more sophisticated promoter behaviours.   Finally, we show the distance between binding sites is highly conserved, thus supporting the idea that bacterial evolution may be partially driven by the physical constraints imposed by the TF search mechanism.

\begin{figure}[t]
\begin{center}
\includegraphics[angle=270,width=\textwidth]{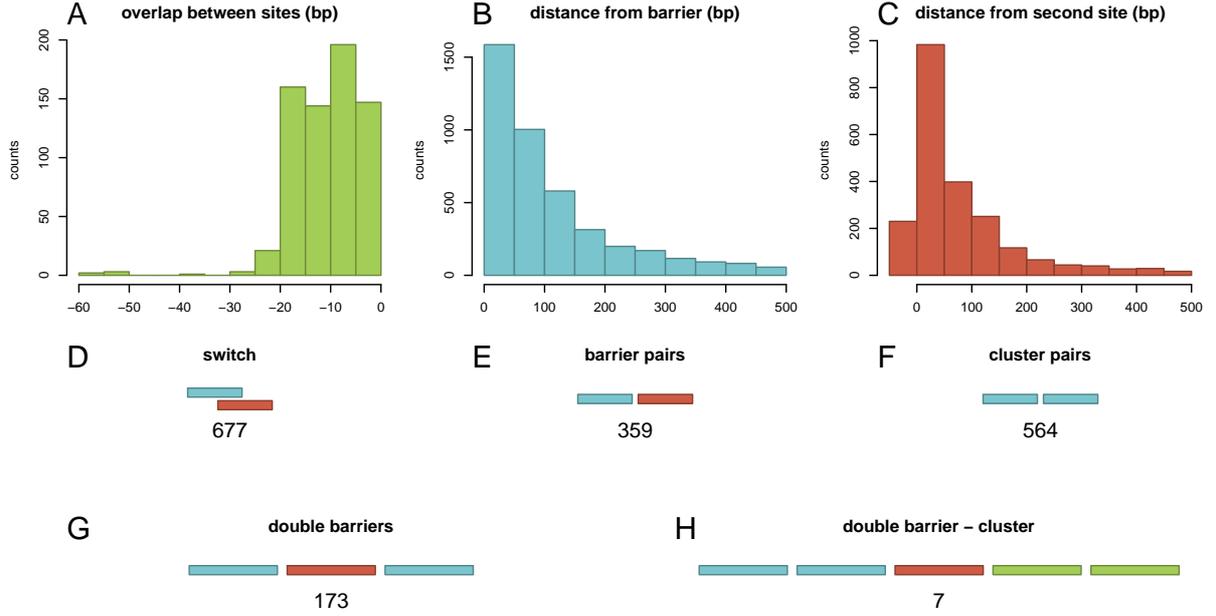}
\end{center}
\caption{\emph{Distribution of promoter architectures in \emph{E. coli}}. We considered the binding sites in \emph{E. coli} K-12, which were listed in RegulonDB \cite{gama_castro_2011}. ($A-C$) plot  histograms of the overlap or the distance between two sites that form a $(A)$ switch, $(B)$ barrier and $(C)$ cluster. Next, we counted the number of  $(D)$ switches, $(E)$ barriers and $(F)$ clusters and found that these building blocks are frequently encountered in \emph{E. coli} K-12 genome. For the barrier and cluster pairs, we consider binding sites that are less than $10\ bp$ apart. In $(G)$ and $(H)$  we presented some examples of complex promoters: $(G)$ double barrier and $(H)$ double barrier cluster. }
\label{fig:motifCounts}
\end{figure}

\section*{Materials and Methods}

\subsection*{GRiP simulations of promoter building blocks}

We used GRiP to simulate the the facilitated diffusion mechanism \cite{zabet_2012_grip,zabet_2012_model}.  GRiP models the diffusion of TFs implicitly via the Chemical Master Equation, as described in \cite{zon_2006}.  Although other facilitated diffusion  simulations incorporate the 3D structure of DNA \cite{lomholt_2009,benichou_2011,brackley_2012,fofanno_2012, bauer_2013}, we do not because TF arrival times in \emph{E. coli} are not significantly dependent on the 3D organisation of DNA \cite{zabet_2013_time}.  The system was parametrised with values estimated from experimental data \cite{zabet_2012_model} and each simulation was run for $3000\ s$, approximately the \emph{E. coli} cell-cyle \cite{rosenfeld_2005}. We used the system-size reduction described in \cite{zabet_2012_subsystem}. The full list of parameters is listed in Table \ref{tab:GRiPTFparams} in the \sm.

\subsection*{fastGRiP simulations of expanded parameter spaces and complex promoters}

Analytical solutions of facilitated diffusion are faster than our GRiP simulations and can provide more insight into the mechanisms underlying the system, but they cannot incorporate real affinity landscapes. We developed a semi-analytical model (which we call fastGRiP) that uses mathematical approximations for the diffusion of TF molecules on non-specific DNA.

Our semi-analytical model uses a continuous time Markov chain, where each state represents a possible promoter configuration.  Each transition represents the propensity of a single binding, unbinding or relocation reaction (in the case of a cluster).  The reaction propensity is equal to: $k=1/t$, where $t$ is the expected time for the reaction to occur.  

The size of the Markov chain is $2^n$, where $n$ is the number of binding sites in the system. This means that the Markov chain grows rapidly with the number of binding sites and, thus, it becomes difficult to solve the system analytically \cite{stewart_1994}, so we use the exact stochastic simulation algorithm \cite{gillespie_1976,gillespie_1977}, which generates a statistically correct trajectory through the Markov Chain  \cite{gillespie_2000}.

\subsubsection*{Binding event}
The propensity of binding is calculated by an adaption of an equation described in \cite{kolesov_2007}
\begin{align}
k_{\textrm{binding}}= \frac{TF_{\textrm{free}}}{\frac{M}{s_l^*}\left(t_r+\overline{t}_{\textrm{association}}\right)}
\end{align}
where $TF_{\textrm{free}}$ represents the number of unbound transcription factor, $M$ represents the length of the DNA segment that is being modeled, $t_r$ represents the time spent during a 1D random walk and $\overline{t}_{\textrm{association}} = 1/\overline{k}_{\textrm{association}}$ represents the time a single TF spends during 3D diffusion, where $\overline{k}_{\textrm{association}}$ is the adjusted association rate to the DNA when assuming a smaller DNA segment; see \cite{zabet_2012_subsystem}.  

When a TF binds randomly to the DNA, it has a $s^*_l/M$ probability of landing within a sliding length of its binding site.  The probability of binding can be expressed as a geometric distribution with expected value of $M/s^*_l$ (the TF is expected to bind after $M/s^*_l$ search attempts).  Each search attempt takes $t_r+ \overline{t}_{\textrm{association}} \ s$ : the time spent during the random walk plus the time spent undergoing 3D diffusion.

If TFs are spaced far apart, $s_l^*$ equals the sliding length, which is approximated by $90\ bp$ based on experimental evidence for \emph{lacI} \cite{elf_2007}.  When a nearby binding site is occupied (the barrier case), $s_l^*$ would represent the size of the reduced region from which a TF could find its binding site during a random walk.  For instance, if there was a barrier $1bp$ away from the binding site, $s_l^*=90/2+1$. This approximation is supported by a recent study \cite{brackley_2013_crowding}, which performed coarse grained molecular dynamics simulations of TFs performing facilitated diffusion and showed that by increasing molecular crowding on the DNA (and, thus, the probability of a barrier forming in the neighbourhood of the binding site) leads to an increase in the number of 1D random walks required to locate the binding site.

When a TF has closely spaced binding sites (the cluster case), and one of the sites is already bound, $s_l^*$ is the same as in the case of a barrier.   Even when neither site is bound, the value of $s_l^*$ must be reduced, because the TF will bind to the first site reached. In the case of two $20\ bp$ length binding sites that are $1bp$ away from each other, $s_l^*=90/2+(20+1)/2$.  

A bound barrier or a bound/unbound cluster would result in a lower probability of a TF randomly landing within a sliding length of its binding site, so the propensity of binding is lower. 

\subsubsection*{Unbinding event}
The propensity of unbinding events can be written as
\begin{align}
k_{\textrm{unbinding}}=\frac{1}{\sum_{i=1}^{s_l^*}\left[\frac{N_{1D}}{s_l^*}\cdot\tau_0\cdot\exp{\left(-\beta E_i\right)}\right]}
\end{align}
where $N_{1D}$ is the number of sliding events in the random walk ($N_{1D}=s_l^2/2$) \cite{wunderlich_2008}, $\tau_0$ is the amount of time spent at the strongest target site, and $\beta E_i$ represents the binding energy at position $i$ in the region over where the random walk takes place \cite{gerland_2002}. 

The size of the region over which the random walk $s^*_l$ is calculated as before, except that $s^*_l$ does not need to be adjusted in the case of unbound clusters, because the neighboring sites do not restrict the random walk.  

When there are barriers, the TF will visit its preferred binding site more often than usual, because its random walk is restricted, thereby increasing the time the TF spends bound. If the second TF in a cluster is unbound, the first TF will sample both binding sites during its random walk and will therefore remain bound much longer.  

\subsubsection*{Relocation event}
The expected time that a molecule moves by 1D random walk from one site to a nearby one, located $d$ nucleotides away, is equal to the expected time of the random walk between the two sites:
\begin{align}
k_{\textrm{relocation}}=\frac{1}{\sum_{i=1}^{d}\left[2\cdot d\cdot\tau_0\cdot\exp{\left(-\beta E_i\right)}\right]}
\end{align}
where $d$ is the distance between the two sites. 

\subsubsection*{Assumptions of fastGRiP}

A main assumption of fastGRiP is that TF binding to non-specific sites can be approximated by an analytical solution. We see that fastGRiP and GRiP provide statistically equivalent outcomes, indicating that this assumption likely holds.  

Other factors that are not simulated explicitly in GRiP might influence protein localization. For instance, fastGRiP does not directly take into account TF-TF interactions, but some of the behaviours of TF-TF interactions can be indirectly included. TFs may influence the binding dynamics via dimerisation or recruitment. If two TFs first dimerise and then bind to the DNA, they can be treated as a single TF in the model, but if two TFs individually bind and then dimerise on the DNA (or if one TF recruits a neighboring TF by influencing the binding affinity of the neighboring site), fastGRiP will not be able to model this behaviour yet, and one should use a comprehensive computational model to simulate the facilitated diffusion mechanism (such as GRiP).  

In addition, TFs may provide steric hindrance to the left and the right of the binding site, and these values may be estimated from DNAse I and MNase footprinting \cite{Fairall1986}. 

Finally, our semi-analytical model (fastGRiP) is just an approximation of a more comprehensive model that considers facilitated diffusion (GRiP) and, thus, it may not capture some of the noise that is generated by the non-uniform landscape and by non-cognate TF molecules.

\section*{Results}

\subsection*{Promoter logic building blocks}

Facilitated diffusion influences the rate at which promoter configurations form by affecting the association and dissociation rates of TFs to/from their target sites. It has been shown both theoretically \cite{ruusala_1992,zabet_2013_time}  and experimentally \cite{hammar_2012} that a TF bound to a strong binding site can form an obstacle that slows the rate of binding of a neighbouring TF. Other studies have suggested that multiple adjacent binding sites for the same transcription factor might enhance TF binding \cite{brackley_2012} and increase gene expression \cite{sharon_2012}.  These experiments and simulations suggest that the spacing of transcription factors help encode transcriptional logic. Here we consider three promoter building blocks in which the spacing between TF binding sites influences the dynamics of TF binding: switches, barriers and clusters.  These three components are found frequently throughout the \emph{E. coli} genome; see Figures \ref{fig:motifCounts}($D-F$).   
 
To investigate the binding of TFs to these three building blocks, we simulated the process by which TFs search for their binding sites using the stochastic simulation framework GRiP \cite{zabet_2012_grip,zabet_2012_model,zabet_2012_subsystem,zabet_2012_review}. The details of the model and parameters are listed in the \emph{Materials and Methods} and in Tables \ref{tab:paramsDescription}, \ref{tab:GRiPTFparams} and \ref{tab:modelTFparams3D} in the \sm.  Please note that some additional results were obtained with an approximation of GRiP called fastGRiP, as detailed later this paper (\emph{Approximating GRiP with fastGRiP}) and in the \emph{Materials and Methods}.

\subsubsection*{Switches}

Many TFs in the \emph{E. coli} genome have overlapping binding sites; Figure \ref{fig:motifCounts}(D).  If two or more TFs have overlapping binding sites, only one of the TFs can bind to that position at a time, resulting in a "switch"-like behaviour.  

\begin{figure*}[t]
\begin{center}
\includegraphics[angle=270,width=\textwidth]{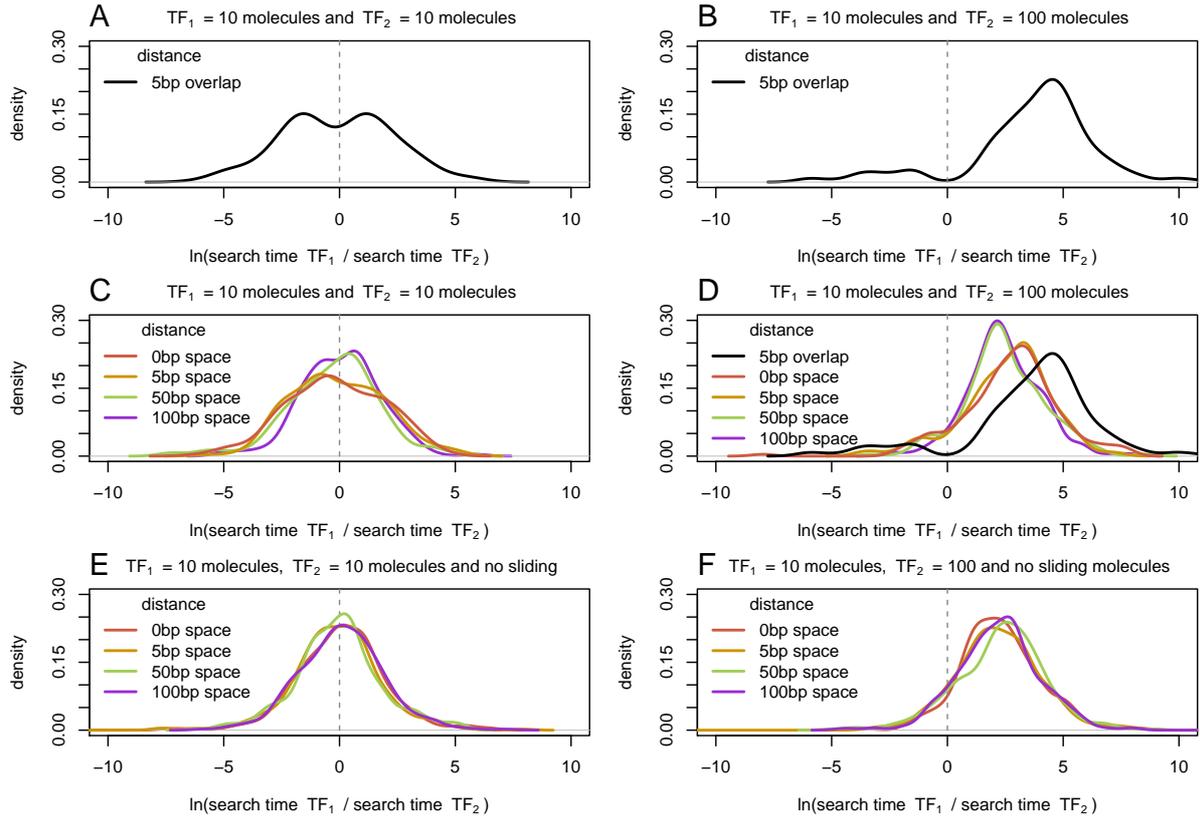}
\end{center}
\caption{\emph{Ratio of TF arrival times for switches and barriers}. Here we show the density plot of the difference in arrival to the two target sites for $(A)$ and $(B)$ switches (overlapping sites by $5\ bp$) and $(C)$, $(D)$, $(E)$ and $(F)$ barriers (distances $\in\{0, 5,50,100\}\ bp$). We considered an overlap of $5\ bp$ since it is the average overlap between two adjacent binding sites; see Figure \ref{fig:motifCounts}$(A)$. We simulated facilitated diffusion in $(A-D)$, but only 3D diffusion in $(E)$ and $(F)$. The set of parameters for the TFs performing facilitated diffusion are listed in Table \ref{tab:GRiPTFparams} in the \sm, while the set of parameters for the system TFs performing only 3D diffusion are listed in  Table \ref{tab:modelTFparams3D} in the \sm. In $(A)$, $(C)$ and $(E)$ the two TF species have the same abundance ($10$ molecules), while in $(B)$, $(D)$ and $(F)$ the second TF is ten times more abundant than the first TF ($TF_1=10$ and $TF_2=100$). Note that in $(D)$ to emphasise the dependence of of the arrival time on the distance we also plot the case of overlapping binding sites.}
\label{fig:switchesBarriersTime}
\end{figure*}

Here, we simulated with GRiP a switch system formed of two overlapping binding sites with two TFs ($TF_1$ and $TF_2$) and we measured the ratio of their respective times to first binding. Figure \ref{fig:switchesBarriersTime}$(A)$ shows that the log ratio of the arrival times ($log(\frac{arrival\ time\ of\ TF_1}{arrival\ time\ of\ TF_2})$) display a bimodal distribution, with a p-value of $0.037$ when performing the dip test \cite{hartigan_1985,diptest_2012}, where each mode of the distribution represents the case of a different TF arriving first.

A particular TF's probability of binding in this competitive environment is also influenced by the differing concentrations and binding affinities of the alternative TFs. TFs with higher concentrations find their binding sites faster than TFs with comparatively lower concentrations (due to a higher number of molecules searching for the binding site); see Figure \ref{fig:switchesBarriersTime}$(B)$. Similarly, a TF with a higher binding affinity would remain bound to the DNA for a longer period of time, thereby preventing other TFs from attaching to its site; see Figure \ref{fig:barrierbuildingblock_heatmap} in the \sm.  These results are also valid in the case of non-uniform affinity landscape; see Figure \ref{fig:GRiPnonuniform} in the \sm.

\subsubsection*{Barriers}

Next, we investigated the influence of adjacent, but non-overlapping  binding sites on promoter configuration.  Previously, it has been shown that the rate of TF binding can be slowed by decreasing the distance between adjacent but non-overlapping TF binding sites  \cite{ruusala_1992,hammar_2012}.  We refer to this interference as the \emph{barrier effect}. 

Even though properties such as DNA bending and electrostatic interactions between TFs could help explain these results,  \cite{hammar_2012} has demonstrated that facilitated diffusion is sufficient to explain the observed barrier effect in the case of a barrier containing \emph{LacI} and \emph{TetR} binding sites.  Previous research found that the TF molecules slide along the DNA maintaining a specific orientation with respect to the DNA (following a helical path) \cite{blainey_2009}. This supports the idea that, in a facilitated diffusion-based model, a TF can find its binding site by 1D diffusion from two directions (by diffusing to the binding site from an upstream or downstream direction). When one of the TFs in a barrier is bound, the other TF can only find its binding site by 1D diffusion from one of these directions, thereby making its arrival less probable and increasing the average time to binding.

Our results support the work of \cite{hammar_2012}; see Figure \ref{fig:switchesBarriersTime}$(D)$. In particular, non-overlapping binding sites that are within half of the sliding length, the binding of the least abundant TFs is slowed down, but not as much as in the case of overlapping binding sites.  When the binding sites are far apart (further than half of the sliding length), the facilitated diffusion mechanism does not significantly influence the rate of TF binding.

To demonstrate that this result is consequence of facilitated diffusion, we compared our standard GRiP simulations to those that only included 3D diffusion. When our model only enabled TFs to find their binding sites via 3D diffusion (no 1D diffusion), the distance between the binding sites had no impact on the TF arrival times; compare Figure  \ref{fig:switchesBarriersTime}$(C)$ to Figure \ref{fig:switchesBarriersTime}$(D)$.  This confirms that the barrier effect was a direct consequence of 1D diffusion along the DNA in our simulations.

Barriers do not significantly affect the total amount of time the TFs spend bound to their binding sites across a cell cycle (the ``total occupancy''); see Figure \ref{fig:barrieroccupancy} in the \sm.  This is the result of two opposing effects: on one hand, the barrier effect increases the average time it takes for a TF to reach its binding site as shown in Figure \ref{fig:switchesBarriersTime}$(D)$ \cite{hammar_2012,zabet_2013_time}, while, on the other hand, once the TF is bound, it stays bound longer by restricting the ability of TF molecules to diffuse away from their sites \  \cite{wang_2012,zabet_2013_time}.  Based on the physics equations we derived for fastGRiP (see \emph{Materials and Methods}), we can demonstrate that although the total occupancy is not significantly different, the rate of TF binding \emph{and} unbinding are reduced in the barrier case; see Figure \ref{fig:barriertransitions} in the \sm.

\subsubsection*{Clusters}

Unlike barriers, whose binding sites are for different TFs, clusters contain multiple binding sites for the same TF.  Therefore, a suitable TF can bind at any site in a cluster.  Bacterial cells frequently have multiple copies of the same binding site clustered together; see Figure \ref{fig:motifCounts}$(F)$ and \cite{hermsen_2006}. Experiments have shown that TF binding site clustering can enhance gene expression \cite{sharon_2012}.

If we consider the facilitated diffusion mechanism, clusters display two opposing behaviours, namely: $(i)$ a TF can slide back and forth between the two nearby binding sites (thus increasing occupancy within that region) and $(ii)$ a bound TF can act as a barrier to the other binding site and a neighbouring empty binding site can also act as a trap, since the TF will attach to the first binding site it reaches, also slowing the rate of binding (thus slowing the rate of binding of other molecules to the second site); see Figure \ref{fig:clusterbuildingblock_heatmap} in the \sm. 

\begin{figure*}[t]
\begin{center}
\includegraphics[angle=270,width=\textwidth]{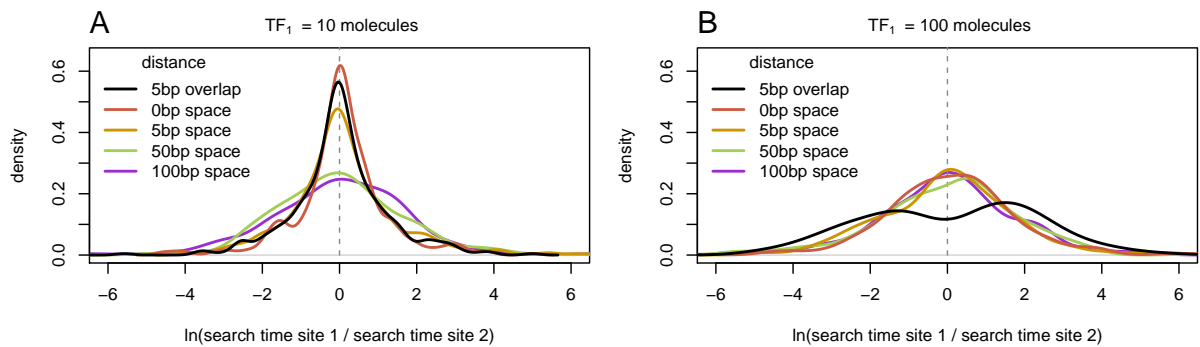}
\end{center}
\caption{\emph{Ratio of TF arrival times in clusters}. We show the density plot of the difference in arrival to the two binding sites (distance between sites $\in\{-5, 0, 5, 50, 100\}\ bp$) of the same transcription factor, $TF_1$.  In $(A)$, $TF_1$ has low abundance ($TF_1=10\ molecules$) and, in $(B)$, $TF_1$ has high abundance ($TF_1=100\ molecules$).}
\label{fig:clustersTime}
\end{figure*}

The balance between these two opposing behaviors depends on the concentration and binding affinity of the binding sites.   For instance, clusters enhance TF binding rates at low concentrations (Figure \ref{fig:clustersTime}$(A)$). However, clusters do not enhance binding rates when the concentration of the TF is sufficiently high (Figure \ref{fig:clustersTime}$(B)$).  In fact, we see an increased degree of bimodality in overlapping bindings sites in a cluster than in a standard switch (with a p-value of $0.025$ when performing the dip test in the cluster case, as opposed to a p-value of $0.037$ in the switch case).

\subsection*{Complex configurations}

Complex promoters are common in the \emph{E. coli} genome: there are $195$ sets of three or more binding sites that are all separated by less than $100\ bp$; see Figure \ref{fig:motifCounts}$(G)$. These complex promoters have diverse structures in the sense that every possible combination of switches, barriers, and clusters that could be constructed from 3, 4, or 5 binding sites will be found in the \emph{E. coli} genome;  Figure \ref{fig:notation} in the \sm\ introduces a nomenclature that we developed to systematically catalogue promoter architectures. An example of switching, barrier and clustered promoters (and combinations thereof) with four different binding sites is shown in Figure \ref{fig:tripletdist} in the \sm. The entire dataset of all promoter architecture classifications is accessible at http://logic.sysbiol.cam.ac.uk/fgrip/db; as introduced in Figure \ref{fig:fivemerdist} in the \sm.

A particular promoter architecture may be enriched in the genome because of evolutionary selection for a certain functional role or as a byproduct of the process by which mutations occur \cite{He2012, Lusk2010}. For instance, clusters can arise from local DNA sequence duplication \cite{Nourmohammad2011}, a common mutation event.  Therefore, similar binding sites tend to be co-located on the genome rather than being interspersed with unrelated motifs (p-value$= 0.019$, chi-squared test).

Interestingly, there is an enrichment for alternating switch-barrier architectures as compared to architectures with the same ratio of switches and barriers (p-value$=8.2\times 10^{-5}$, chi-squared test; see Figure \ref{fig:fivemerdist} in the \sm) and this architecture is enriched for genes activated by nitrite/nitrate (p-value$=4.4\times 10^{-13}$, binomial probability distribution), suggesting that this architecture might play an important role. In order to understand possible functional roles of complex promoter architectures, we simulated the TF search process of a number of these complex promoters.
 
\subsubsection*{Approximating GRiP with fastGRiP}
Fully stochastic simulations (such as GRiP) are too computationally intensive to simulate complex systems. Analytical solutions of facilitated diffusion are faster and provide more insight into the mechanisms underlying the system, but they cannot incorporate real affinity landscapes. Therefore, we developed a semi-analytical model (fastGRiP), which is based on the behaviour of the three building blocks (switches, barriers and clusters).  fastGRiP uses a continuous time Markov chain, where each state represents a possible promoter configuration and each transition represents a single binding, unbinding or relocation event (the case of a TF jumping between two adjacent TF binding sites in a cluster).  The full description of fastGRiP and the associated equations are presented in \emph{Materials and Methods}.  It is compared to GRiP in Figure \ref{fig:comparisonToGRiP} and the runtime is analysed in Figure \ref{fig:runtime} in the \sm.

\subsubsection*{Two-Sided Barriers}

The first complex promoter that we investigated is the \emph{ABA} pattern, which has two identical sites that surround another site. Each adjacent pair of TF binding sites forms a barrier, and the two binding sites of the same type can form a cluster if the binding sites are close enough to one another; see Figure \ref{fig:motifCounts}$(G)$. 
 
In the building block section, we described how a bound TF slows the rate of binding to an adjacent site via the barrier effect.   We wished to see how much this barrier effect would be amplified if barriers surrounded a TF binding site on both sides.  Therefore, we focused our analysis on how the ABA pattern affected the ability of all three TFs to be bound at once, what we call the \emph{AND configuration}.  
 
\begin{figure*}[t]
\begin{center}
\includegraphics[angle=270,width=\textwidth]{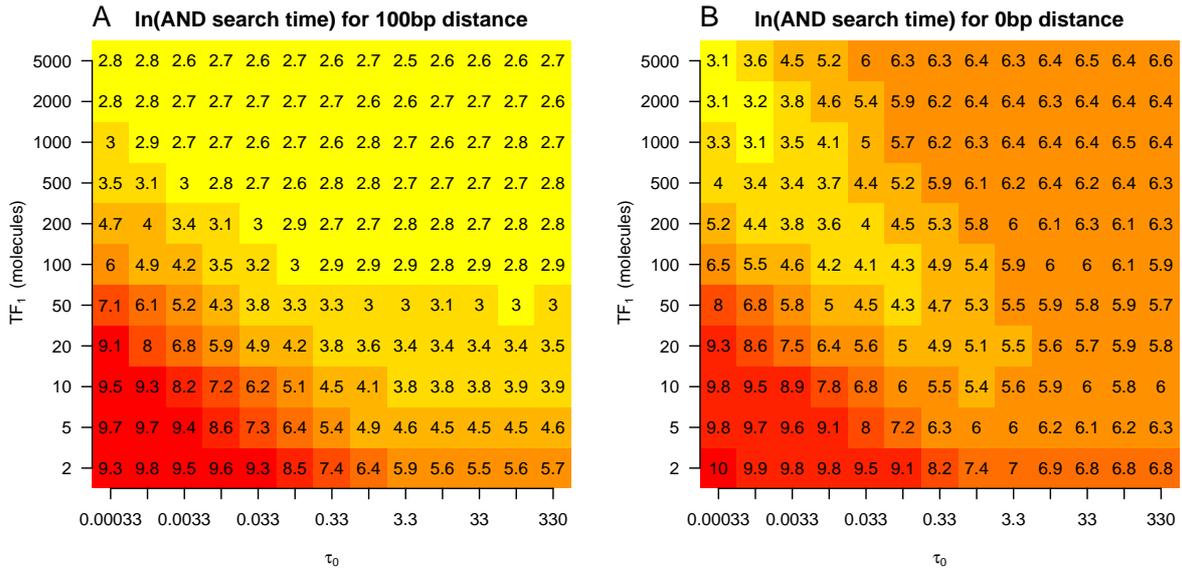}
\end{center}
\caption{\emph{Heatmap of the ln of the first time all three sites are occupied in double sided barriers}. We considered the promoter configuration ABA, where A is the target site of $TF_1$ and B is the target site of $TF_2$. The system consists of $10$ molecules of  $TF_2$ with a binding affinity scaling parameter of  $\tau_0=0.33$; see \emph{Materials and Methods}. The distance between adjacent binding sites is $(A)$ $100\ bp$ and $(B)$ $0\ bp$. We vary the abundance and DNA binding affinity of $TF_1$.}
\label{fig:doubleBarriersTimeHeatmap}
\end{figure*}

When the binding sites were $100\ bp$ apart (far enough \emph{not} to be influenced by the barrier effect), the simulations predicted that increasing the concentration and binding affinity of the TF binding to the outer binding sites would increase the likelihood that all three sites are occupied simultaneously (the \emph{AND configuration});  see Figure \ref{fig:doubleBarriersTimeHeatmap}$(A)$. Since the binding sites are far enough away as to not be influenced by the barrier effect, these results are consistent with the expected outcome of a thermodynamic ensemble model.

In contrast, when the binding sites are close together ($0\ bp$), we observed a behaviour that cannot be explained within the thermodynamic framework. More specifically, at high concentrations and binding affinities, the \emph{AND configuration} forms slowly since the outer two binding sites are frequently bound, restricting the binding of the central binding site (through the barrier effect).  Furthermore,  when both the concentration and binding affinity of the outer sites are low, the \emph{AND configuration} forms slowly, because the outer sites are unlikely to be occupied concurrently.  The case in which the \emph{AND configuration} forms fastest is when the outer binding sites have a low binding affinity and high abundance; see Figure \ref{fig:doubleBarriersTimeHeatmap}$(B)$. 

These results also illustrate the magnitude to which the arrival times of TFs can be affected by facilitated diffusion. For instance, when the TFs that bind to the outer two binding sites had an abundance of $2$ molecules and a $\tau_0$ of $0.33$, the AND configuration formed in approximately $1600\ seconds$ ($e^{7.4}$) in the case of far away TF binding sites, and in approximately $3600\ seconds$ ($e^{8.2}$) in the case of closely spaced TF binding sites. Therefore, on average, the barrier effect delayed the formation of the AND configuration by $2000\ seconds$ in this scenario. Note that, while in the former case, the average time to reach the AND configuration is half of the \emph{E. coli} cell cycle (the cell cycle is about $3000\ seconds$), in the latter case, this time is longer than the average length of the cell cycle. This example shows that the spacing between binding sites can influence the timing of TF binding events at time scales, which would be biologically relevant. In fact, we found that the percentage of simulations where the AND configuration is reached within half of cell cycle is dependent not only on the number of binding sites in the promoter, but also on the distance between the sites; see Figure \ref{fig:biologicallyrelevant} in the \sm.

Next, we investigated whether this model could provide insight into genome organisation. In the \emph{E. coli} genome, there are several hundred triplets of closely spaced binding sites, including \emph{pdhR, dpiBA} and \emph{moeAB}.  Among these triplets that are less than $50bp$ apart, the outer binding sites are more likely to have lower binding affinities than the central binding site (p-value$=0.0018$, chi-squared test; see Figure \ref{fig:hillvalleybio} in the \sm). In addition, there is strong anti-correlation (p-value$=2.9\cdot 10^{-5}$, Fisher exact test) between having stronger binding sites and having higher TF concentrations, as measured by APEX \cite{Lu2007}, in the TFs binding to the outer binding sites compared to the TF binding to the inner binding site; see Figure \ref{fig:hillvalleybio} in the \sm. This suggests that the organisation of promoters in \emph{E. coli} may be optimised for having multiple TFs bound at once (instead of two or one) or optimised for allowing binding to the middle binding site.  The \emph{AND configuration} has been shown to play an important role in \emph{E. coli}. Cox III \emph{et al.} \cite{cox_2007} constructed $288$ synthetic promoters in \emph{E. coli} and found that the preferred transcriptional mode between three adjacent sites is the AND logic.

\subsubsection*{Double barrier-cluster}

Next, we considered the case of a site being flanked by two identical clusters (the \emph{AABAA} pattern) with $0\ bp$ between each adjacent pair of TFs.  We wished to determine whether combining clusters and barriers would produce a promoter logic pattern that could not be explained by the behaviour of clusters or barriers alone.  We compared the behavior of the \emph{AABAA} pattern to similar scenarios containing only barriers (five different adjacent TFs, an \emph{ABCDE} pattern) and only clusters (two pairs of clusters separated from a central TF by $100bp$, an \emph{AA-B-AA} pattern).  

\begin{figure*}[t]
\begin{center}
\includegraphics[angle=270,width=\textwidth]{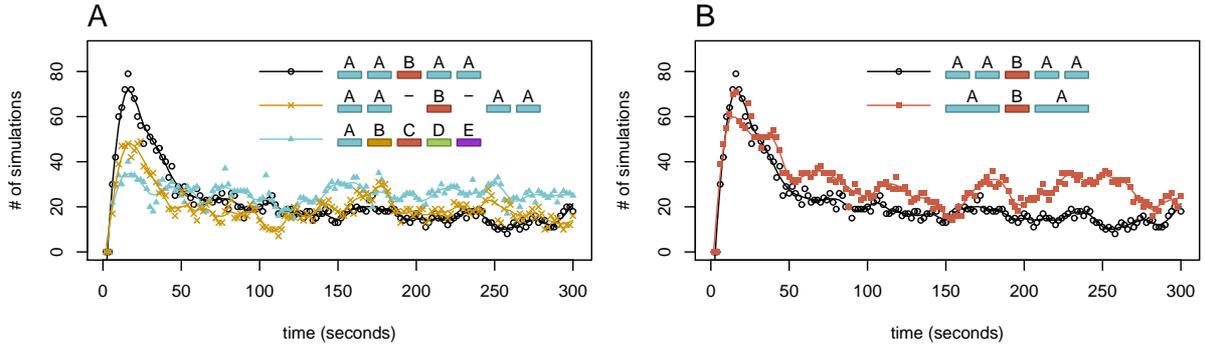}
\end{center}
\caption{\emph{Impulse behavior of \emph{AABAA}}. We show the number of simulations (out of 400) have only the central binding site bound, over time, for the AABAA configuration compared to  $(A)$ AA-B-AA and ABCDE and $(B)$ ABA. For example, in the AABAA scenario, the y-axis represents the number of simulations out of the $400$ in which the B site is bound and none of the A sites are bound. Note that we start the simulation with ``naked'' DNA (no TFs bound). }
\label{fig:AABAApulses}
\end{figure*}

In the AABAA scenario, when we graph the number of simulations in which only the central TF is bound, over time (starting from DNA that has no TFs bound), we observe an impulse behaviour-- a short period of time in which there is a higher probability of this configuration occurring than observed at equilibrium; see Figure \ref{fig:AABAApulses}$(A)$.  Alternatively, the graph can also be read as the number of cells in a population in which only the central TF is bound after x seconds. If we were to consider $TF_B$ to be an activator and $TF_A$ to be a repressor, an impulse could possibly result in a short burst of gene expression.

The impulse behavior is influenced by (i) how often the central TF binds first (ii) how long this configuration lasts before other TFs come and bind to the DNA. In the AABAA configuration, $TF_B$ (binding to site B) has a relatively high rate of binding, because the A binding sites (the clusters) will act as obstacles to one another since a $TF_A$ will bind to the first A binding site it encounters. This configuration will also remain a relatively long time, because the central bound TF ($TF_B$) acts as a barrier that slows the rate of binding of the other TFs.  

When there are no clusters (ABCDE), all the TFs have an equal probability of binding first.  When there are no barriers (AA-B-AA), $TF_B$ cannot act as a barrier to slow the binding of other TFs, so alternate configurations are more rapidly assumed since the binding of other TFs is not obstructed. Therefore, in both the ABCDE and AA-B-AA configurations the size of the impulse is reduced. This illustrates that the combination of barriers and clusters can result in different responses than each component independently; see Figure \ref{fig:mechanism} in the \sm.

We wondered whether the \emph{AABAA} pattern acted similarly to the double barrier (\emph{ABA}) described in the previous section.  The presence of a cluster would increase the time a TF is bound to the DNA and the rate of TF binding; however, we were curious if additional factors beyond these two also contributed to the logic of the AABAA promoter.  We simulated a double barrier \emph{ABA} pattern with the outer two TFs having twice the length and twice the binding strength.  This scenario displayed similar impulse behaviour, but the results displayed more stochasticity than the \emph{AABAA} pattern ($\sigma^2_{AABAA} = 17.3$ and $\sigma^2_{ABA} = 31.3$, while the Fano factor for the AABAA configuration is $1.02$ and for the ABA configuration is $1.15$; F-test: p-value $= 2.2e-16$); see Figure \ref{fig:AABAApulses}$(B)$.  This is supported by recent single cell imaging study that suggests that increasing the strength of a binding site could increase transcriptional noise \cite{Dadiani2013}.

These results indicate that the combination of clusters and barriers can create qualitatively different behaviors than their individual components and that promoter organisations influence the stability of specific promoter configurations.

\subsubsection*{Evolution of complex promoters in \emph{E. coli}}

Because the behaviours of barriers and clusters depends on the spacing of binding sites, one would expect there to be evolutionary selection pressure keeping binding site spacing conserved during the evolution of promoters in \emph{E. coli}.

We compared the insertion-deletion (indel) rates and the single base pair substitution rates of different regions of promoters.  The evolutionary events were parsed into the following categories: between transcription start site (TSS) and first binding site, within binding sites, between closely spaced binding sites, between binding sites farther than $100\ bp$ apart, and between the last binding site and the termination sequence. The indel rates and base pair substitution rates were calculated in a pair-wise fashion between \emph{E. coli} K-12 (the main \emph{E. coli} reference genome) and the other five  NCBI-designated reference \emph{E. coli} genomes ( O157:H7, IAI39, UMN026, O83:H1 and O104:H4). Note that we control for the different DNA sequence lengths in each category since our mutation rates are defined as:  $\frac{number\ of\ mutation\ events}{length\ of\ sequence}$.

\begin{figure*}[t]
\begin{center}
\includegraphics[angle=270,width=\textwidth]{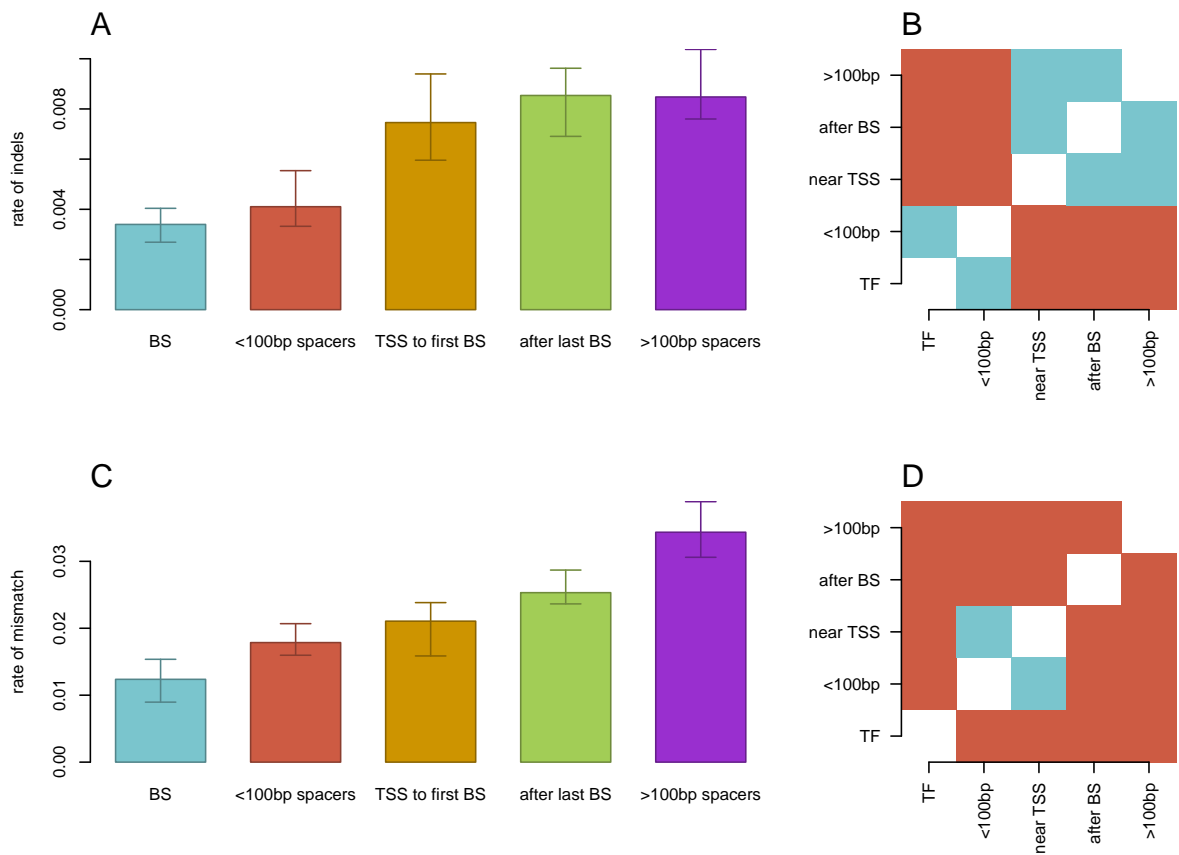}
\end{center}
\caption{\emph{Differential evolution of promoter components in \emph{E. coli}}. We compared $(A)$  and $(B)$ indel rates and $(C)$  and $(D)$  SNP rates between \emph{E. coli} K-12 and other reference strains across different promoter regions. $(B)$ and $(D)$  display results of a paired-T-test with Holm corrections; significantly different promoter regions are red and insignificant pairs are blue (threshold: p-value $=0.05$). We considered the following five cases: $(A)$ within binding sites (BSs), $(B)$ between closely spaced binding sites ($<100$ bp spacers), $(C)$ between binding sites farther than $100\ bp$ ($>100$ bp spacers), $(D)$  between TSS and first binding site (TSS to first BS) and $(E)$ between the last binding site and the termination sequence (after last BS).}
\label{fig:evolution}
\end{figure*}

Figure \ref{fig:evolution} shows that regions between closely spaced TFs ($<100bp$ apart) had similar indel rates as TF binding sites, but had significantly higher rates of single base pair substitutions.   In contrast, regions between distant TF binding sites ($>100bp$ apart, and therefore not influenced by facilitated diffusion) had much higher rates of indels and mismatches.  These results indicate that regions between closely spaced TF binding sites are conserved in terms of length, but not as highly conserved in terms of sequence.  We see that across all closely spaced TF binding sites in all six strains, only two TF binding sites change their relative distance by more than 1bp.  This suggests that there may be evolutionary selection pressure to keep the distances between TF binding sites conserved. Although this evolutionary analysis does not provide direct evidence that facilitated diffusion plays a functional role, these results are compatible with our proposition that TF binding site spacing may have a facilitated-diffusion driven functional role.

Nevertheless, binding site spacing may also be conserved for other reasons. For instance, the distance between adjacent binding sites within promoters are conserved in order to preserve the distance between the binding sites and the TSS (in order to conserve the effect of the binding site on the gene regulation). To investigate this assumption, we also looked at the conservation of the distance between the first binding site in the \emph{cis-}regulaory region and the TSS and found that these regions are not conserved in both distance and sequence; see Figure \ref{fig:evolution}. This indicates that the conservation of the distance between binding sites is not influenced by the distance between the binding sites and the TSS. 

Additionally, TFs that display direct interactions with other TFs or act via DNA-bending may have certain binding site spacing requirements.  If this was the primary reason for the conservation of TF spacing, the spacing of TF binding sites that act via these mechanisms would be more conserved than those that do not.  However, we see that the spacing between \emph{all} closely packed TFs are significantly conserved, suggesting that either \emph{most} closely spaced TF binding site pairs participate in TF-TF interaction or other mechanisms (such as facilitated diffusion) could also influence the selection pressure acting on preserving the distance between binding sites.

\section*{Discussion}

The prediction of TF binding to promoters has received significant attention in the literature, as it is the first step towards developing mechanistic models of gene expression.  Transcriptional logic is often assumed to be independent of the spacing of TF binding sites and only associated with which TFs can bind to the promoter.  The underlying principle behind this assumption is that the cell operates as a well-stirred reactor, which can lead to misleading results because of rapid TF-DNA rebindings \cite{zon_2006}.

Alternatively, the binding of TFs has been predicted by scanning the DNA for a PWM and then calculating the probability of binding using a statistical thermodynamic framework to take into account TF concentration.  \cite{ackers_1982,djordjevic_2003,bintu_2005_model,roider_2007,chu_2009,zhao_2009} and steric hindrance on the DNA \cite{foat_2006,sadka_2009,wasson_2009,hoffman_2010,simicevic_2013}. However, these models assume that TFs are bound at thermodynamic equilibrium, even though thermodynamic equilibrium might not be reached in the time frame of a cell cycle. 

Within a facilitated diffusion context, the distance between TF binding sites on the DNA could potentially encode for transcriptional logic in a way that the classical thermodynamic models cannot capture \cite{ruusala_1992}. In fact, experimental studies have shown that when binding site spacing is manipulated, the occupancy of the site is affected \cite{hammar_2012} and this can even influence transcription \cite{cox_2007,khoueiry_2010,sharon_2012}.  In this paper, we present a theoretical explanation of how binding site spacing could encode facilitated diffusion-based transcriptional logic.

\subsection*{Combining promoter logic building blocks to form complex promoters}

We identified three examples of how spacing between binding sites can influence the dynamics of TF binding, namely: switches, barriers, and clusters. 

Previous studies have suggested that switches influence transcriptional logic of prokaryotic promoters \cite{hermsen_2006}; see Figure \ref{fig:switchesBarriersTime}$(A)$ and  Figure \ref{fig:switchesBarriersTime}$(B)$. \emph{In vivo} experimental studies show that binding site occupancy depends on the distances between TF binding sites, due to the \emph{barrier effect} \cite{hammar_2012} and this change in occupancy can affect gene transcription \cite{cox_2007,khoueiry_2010,sharon_2012}; see Figure \ref{fig:switchesBarriersTime}($C-F$). Finally, we found that under certain TF abundances and affinities, when two identical sites are close (clusters) the difference in association rates to the two sites is reduced, as suggested in \cite{janga_2008}; see Figure \ref{fig:clustersTime}.  These three building blocks are found frequently in the \emph{E. coli} genome; see Figure \ref{fig:motifCounts}($D-F$).  

Although others have suggested that co-localisation of TF binding sites can influence the dynamics of TF binding, none of these studies have analysed complex promoters that combine barriers, switches and clusters, which can encode complex behaviors.  Describing promoters in terms of these building blocks gives us a language to help classify complex promoters by their structures and look for patterns in promoter organisation.  Our analysis also indicates that taking into account the distribution of promoter building blocks may be a useful lens for evaluating the evolution of promoters.

\subsection*{Complex promoters}

Two of the complex configurations that we studied in depth were the double barrier (an ABA configuration) and the double barrier-cluster (an AABAA configuration).  In the double barrier scenario, we observed that the time for all TFs to find their binding sites depended on the concentration and binding affinity of the TFs.  The promoter organisation was optimised for having all three TFs bound at once when the outer TFs had higher concentration and lower binding affinity than the central TF; see Figure \ref{fig:doubleBarriersTimeHeatmap}.  Interestingly, this organisation was significantly enriched within the \emph{E. coli} genome.  

In the double barrier-cluster scenario, the promoter configuration can display a temporal impulse in the occupancy of the middle site. A similar type of response can also be produced by an incoherent feed-forward loop in the gene regulatory network \cite{shen_orr_2002}. On one hand, the advantage of having the impulse response encoded in the occupancy of the promoter (and not in the gene regulatory network) is the fact that the response is faster and the metabolic cost is lower (the gene expression process is both slow and metabolically expensive for the cell) \cite{zabet_2009}. On the other hand, the disadvantage of having the impulse response encoded in the occupancy of the promoter is that it becomes difficult to sustain the impulse for longer time intervals and this is where the gene regulatory network overcomes the limitation of the promoter occupancy. Hence, depending on the system requirements, i.e. biological context, the temporal impulse response can be encoded in the promoter configuration (faster response) or in the gene regulatory network (longer impulse). 

\subsection*{Facilitated diffusion as a lens for interpreting experimental data}

Frequently there is an assumption that two TFs with correlated binding behaviours must interact directly \cite{hermsen_2006,sadka_2009,kaplan_2011}. Here, we show that TFs can influence the binding of neighbouring sites without direct interaction. We agree that protein-protein interactions are important for determining transcriptional logic \cite{he_2009,cheng_2013}, and in many cases OR, XOR, and AND logic could be primarily encoded in these interactions; however, screens for these interactions should consider a facilitated diffusion based model as their null hypothesis.

For example, Cheng \emph{et al.} \cite{cheng_2013} used a statistical thermodynamics model to analyse ChIP data and identified that the data is best explained when including blocking of binding (antagonistic effects) even in the case when the sites do not overlap. Interestingly, they found a bias in the distance between the non-overlapping sites of up to $30\ bp$, which suggest a possible barrier effect being involved. We are not claiming that the facilitated diffusion is the only possible explanation for the observed behaviour, but rather that this might be one possible explanation for the observed results. 

The possible functional role of TF spacing also opens up interesting questions in an evolutionary context: are the locations of TF binding sites influenced by the physics of diffusion?  We compared the indel rates in six \emph{E. coli} strains (K12, O157:H7, IAI39, UMN026, O83:H1 and O104:H4) and our results showed high conservation of the spacing between binding sites for spaces smaller than $100\ bp$ (similar with the conservation of the binding sites themselves); see Figure \ref{fig:evolution}. In contrast, the DNA sequence of the spaces between the binding sites is not conserved and neither is the distance between the binding sites and the TSS. Put together, these results suggest that the evolution of bacterial systems might be influenced by the facilitated diffusion mechanism.

\subsection*{Computational tool}

To aid biologists in analysing complex promoter behaviours, we provide a semi-analytical model (called fastGRiP) through an intuitive web interface (http://logic.sysbiol.cam.ac.uk/fgrip/; also see Figure \ref{fig:fastGRiP} in the \sm), which leads to only negligible deviations from the full model of facilitated diffusion (GRiP).  It is significantly faster than GRiP and allows investigations of complex promoters under a wider set of parameters within short simulation times; see Figure \ref{fig:runtime} in the \sm.  Furthermore, our complete classification of \emph{E. coli} promoters can be browsed at http://logic.sysbiol.cam.ac.uk/fgrip/db; with the option to download the dataset for further analysis.

\subsection*{Testing the proposed model with experiments}

Our model predictions may be tested by constructing specific synthetic promoters and measuring TF binding kinetics (e.g. via \emph{in vivo} single molecule microscopy experiments for low abundance TFs) \cite{elf_2007,hammar_2012}  and gene expression (via qPCR or luciferase assays).  The parameters that one would wish to manipulate in these experiments include: $(i)$ the distance between binding sites (varied between 0 bp and 100 bp), by synthesizing different promoter sequences $(ii)$ the abundance of the TF, by using an inducible promoter to control the expression of the TF and $(iii)$ the binding affinity of the TF to its binding site, which can be somewhat controlled by manipulating the DNA sequence of the binding motif or adjusting the salt concentration.  Note that TF abundance and binding affinity can only be roughly adjusted, so only qualitative comparisons can be made.  

For instance, to test whether the double barrier cluster scenario (AABAA) can generate a noticeable impulse behavior in terms of gene expression, one needs to be synthesize a promoter where the TF that binds to the middle site is an activator (e.g. CRP) and the TFs that bind to the surrounding clusters are repressors (e.g. lacI).  Both TFs must be inducible, so that they can both be turned on at similar times, and a luciferase assay could be used to measure gene expression over time. We expect that the double barrier cluster scenario generates an impulse in gene expression, but that this would not be the case if the clusters are far apart from the central TF.

Biophysicists have studied facilitated diffusion for over thirty years, but the focus has been on understanding the fundamental properties of this mechanism. Here we demonstrate that facilitated diffusion could influence the transcriptional logic of common \emph{E. coli} promoter architectures and that these architectures are highly conserved between strains.  We provide a framework, in terms of switches, barriers and clusters, for classifying promoter architectures by their facilitated-diffusion based transcriptional logic and provide a web service to allow biologists to easily analyse the TF binding dynamics of bacterial promoters.  We hope that this is a first step towards bridging between the facilitated diffusion and gene regulation research communities.

\section*{Acknowledgements}

We would like to thank Mark Calleja for his support with configuring our simulations to run on CamGrid, Robert Stojnic  for useful discussions and comments on the manuscript, and Max Hodak for providing advice about our choice of web framework and for commenting on the manuscript. 

\emph{Funding}: This work was supported by a Marshall Scholarship [to DE]; Medical Research Council Bioinformatics Training Fellowship [G1002110 to NRZ] and Royal Society University Research Fellowship [to BA].

\appendix


\begin{figure}[h]
\centering
\includegraphics[angle=270,width=\textwidth]{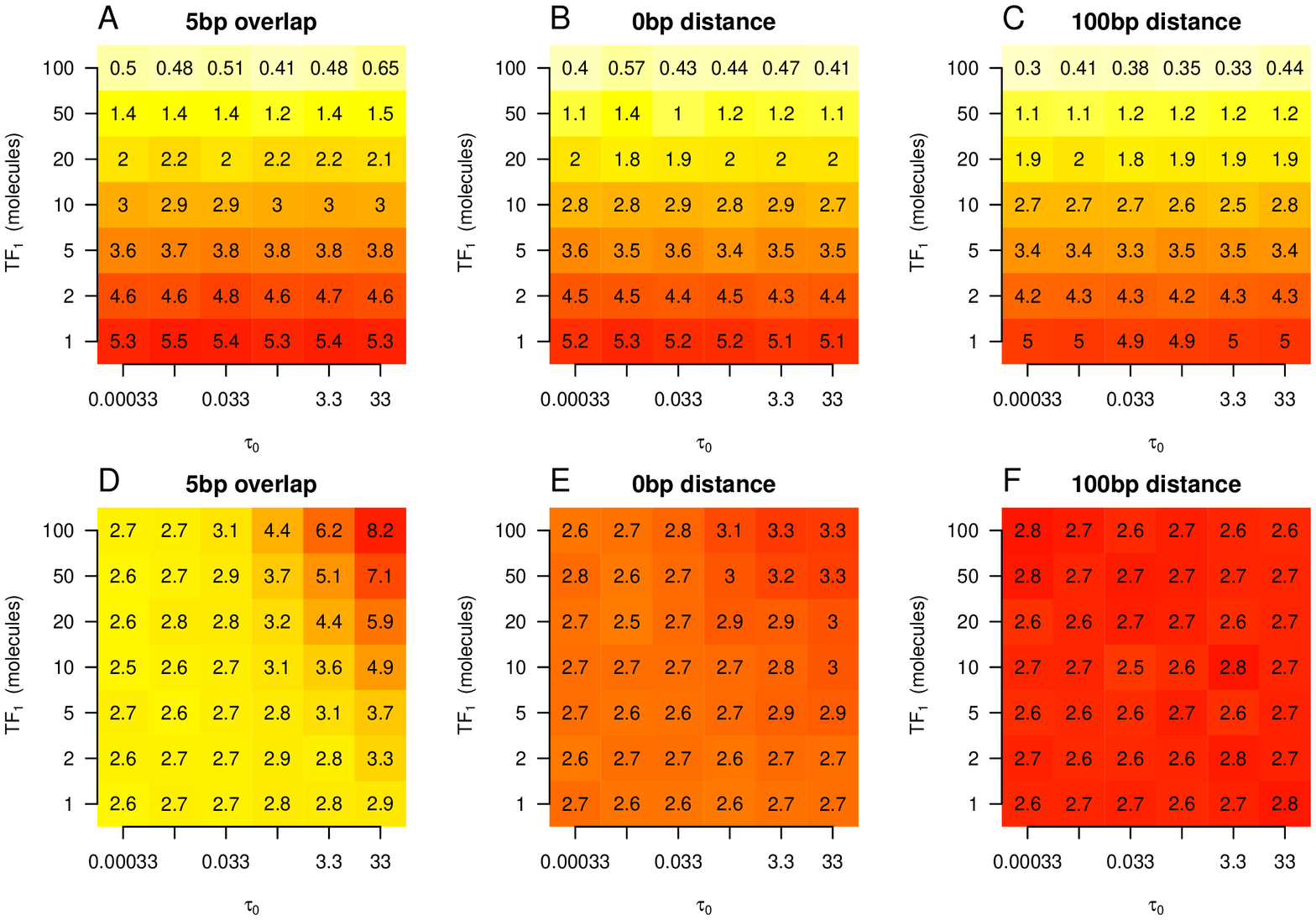}
\caption{\emph{TF arrival time in fastGRiP for asymmetric system}. We simulated the effects of co-localisation of binding sites for two different types of TFs, which we will refer to as $TF_1$ and $TF_2$.  For \onel\ and \fourl\ switches (5 bp overlap); \twol\ and \fivel\  barriers ($0\ bp$ distance between sites) and \threel\ and \sixl\ far apart binding sites ($100\ bp$ distance between sites). We varied the binding affinity parameter ($\tau_0$) and abundance of $TF_1$ and kept the parameters of $TF_2$ constant at $\tau_0=0.33$ and 10 molecules. ($A-C$) show the natural log of the arrival time (the time it takes for $TF_1$ to first reach its binding site in ln(seconds)) for $TF_1$ . ($D-F$) show the natural log of the arrival time for $TF_2$. These values represent the average ln(arrival times) across $400$ simulations.   ($A-C$) We see that in all cases (switch, barrier, far away),  $TF_1$ binds faster at higher abundances, while the binding affinity has no effect on the arrival time of $TF_1$.  We see that binding of $TF_1$ is slightly slower in switches, and to a lesser extent in barriers, compared to the case of all binding sites being far apart.  As we see in \sixl, changing the parameters of $TF_1$ have no effect on the binding of $TF_2$ when the binding sites are far apart.  This supposition of positional independence is a key component of most thermodynamical models.  However, we show that in the case of switches and barriers, increasing the abundance or binding affinity of $TF_1$ increases the arrival time for $TF_2$.} \label{fig:barrierbuildingblock_heatmap}
\end{figure}

\clearpage
\begin{figure}
\centering
\includegraphics[angle=270,width=\textwidth]{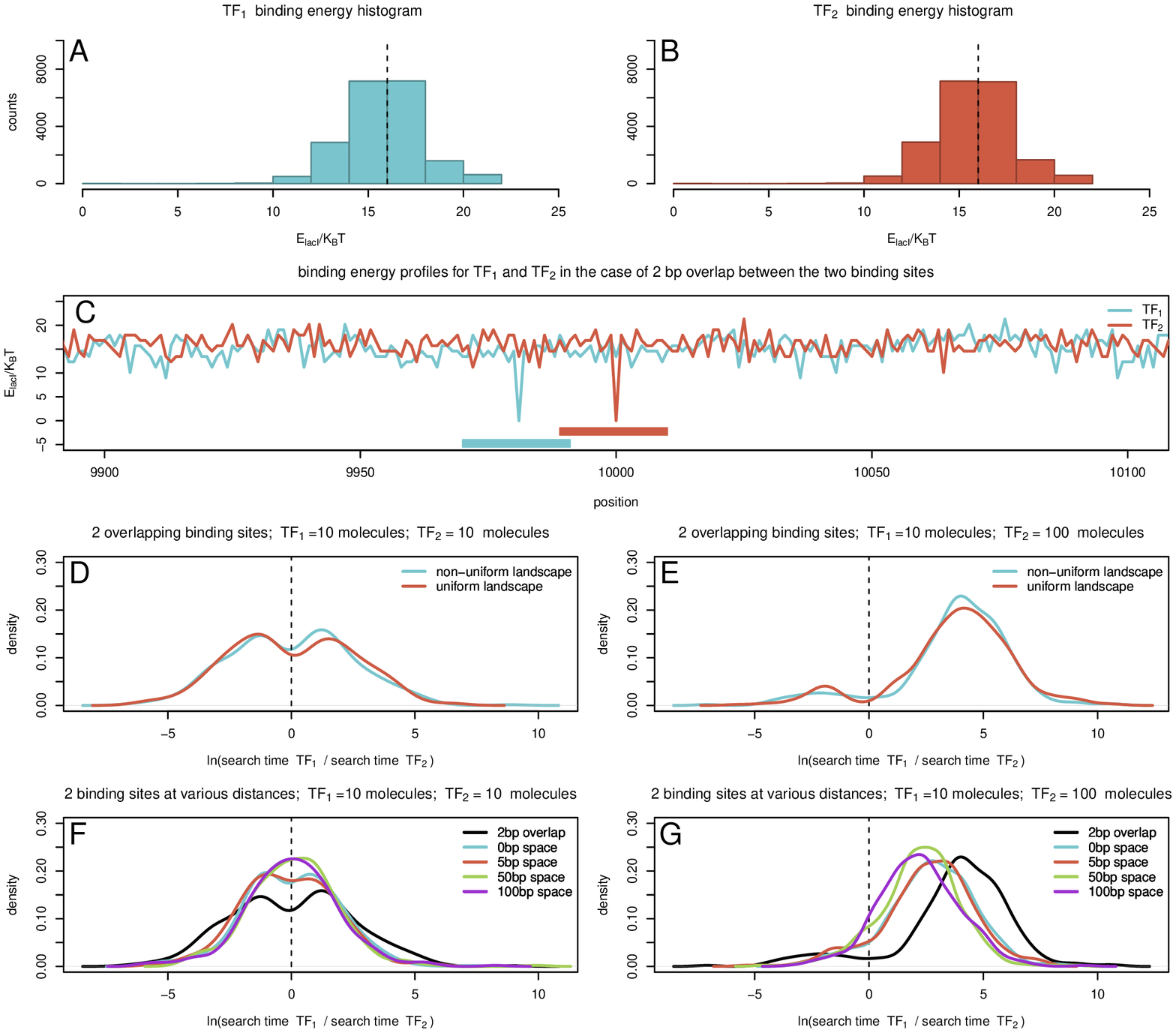}
\caption{\emph{Facilitated diffusion on a non-uniform landscape}. For the non-uniform landscape, we considered the case of a random $20\ Kbp$ DNA sequence and the two binding sites are located in the middle of the sequence; see \threel. The full list of parameters for the non-uniform landscape is listed in Table \ref{tab:GRiPnonuniform}. \onel\ and \twol\ Binding energy histogram for the non-uniform landscape for $TF_1$ \onel\ and $TF_2$ \twol. The mean binding energy is $16\ K_BT$, which is the same as in the case of uniform landscape. \threel\ We also plotted the binding energy profile for the non-uniform landscape case within the region $9900..10100$ of the randomly generated landscape. \fourl\ and \fivel  \emph{Ratio of TF arrival times for switches }. We performed a set of $X=200$ simulations for each set of parameters in the case of the uniform landscape and $X=400$ in the case of non-uniform landscape. The two sites for the two TFs overlap by $5\ bp$ in the case of the uniform landscape and $1\ bp$ in the case of the non-uniform landscape. The two TFs have the same abundance ($TF_1=TF_2=10$) in \fourl\ and different abundances ($TF_1=10$ and $TF_2=100$) in \fivel. \fourl\ For a non-uniform landscape, we found that the $TF_2$ gets first to its target site in $51.25\%$ while $TF_1$ in $48.75\%$ of the cases when both TFs have the same set of parameters, thus, creating an slightly unbalanced system. When computing the waiting time within one sliding length of the target site, We found that $TF_2$ is on average  $4.2\%$ faster in the 1D movement compared to $TF_1$. Thus, in a symmetric system, the TF that moves faster within half of sliding length from the target site can reach its target site first. This result contradicts the prediction that slowing down the TF near the target site (also known as the  funnel effect) reduces the search time \cite{weindl_2007,weindl_2009,brackley_2012}, but, in those studies, the authors considered a biased random walk and did not consider multiple sites and multiple DNA binding proteins. \sixl\ and \sevenl\ \emph{Ratio of TF arrival times for switches and barriers}.  We performed a set of $X=400$ simulations for each set of parameters. The two sites for the two TFs overlap by $1\ bp$ or are separated by: $0\ bp$, $5\ bp$, $50\ bp$ and $100\ bp$. \sevenl\ When $TF_2$ has a higher abundance than $TF_1$ ($TF_1=10$ and $TF_2=100$), then decreasing the distance between the two sites accentuates the difference between the arrival time at the two sites. These results are similar with the case of uniform landscape, except that the distributions are not as separated as in the case of uniform landscape.} \label{fig:GRiPnonuniform}
\end{figure}

\clearpage
\begin{figure}
\centering
\includegraphics[angle=270,width=\textwidth]{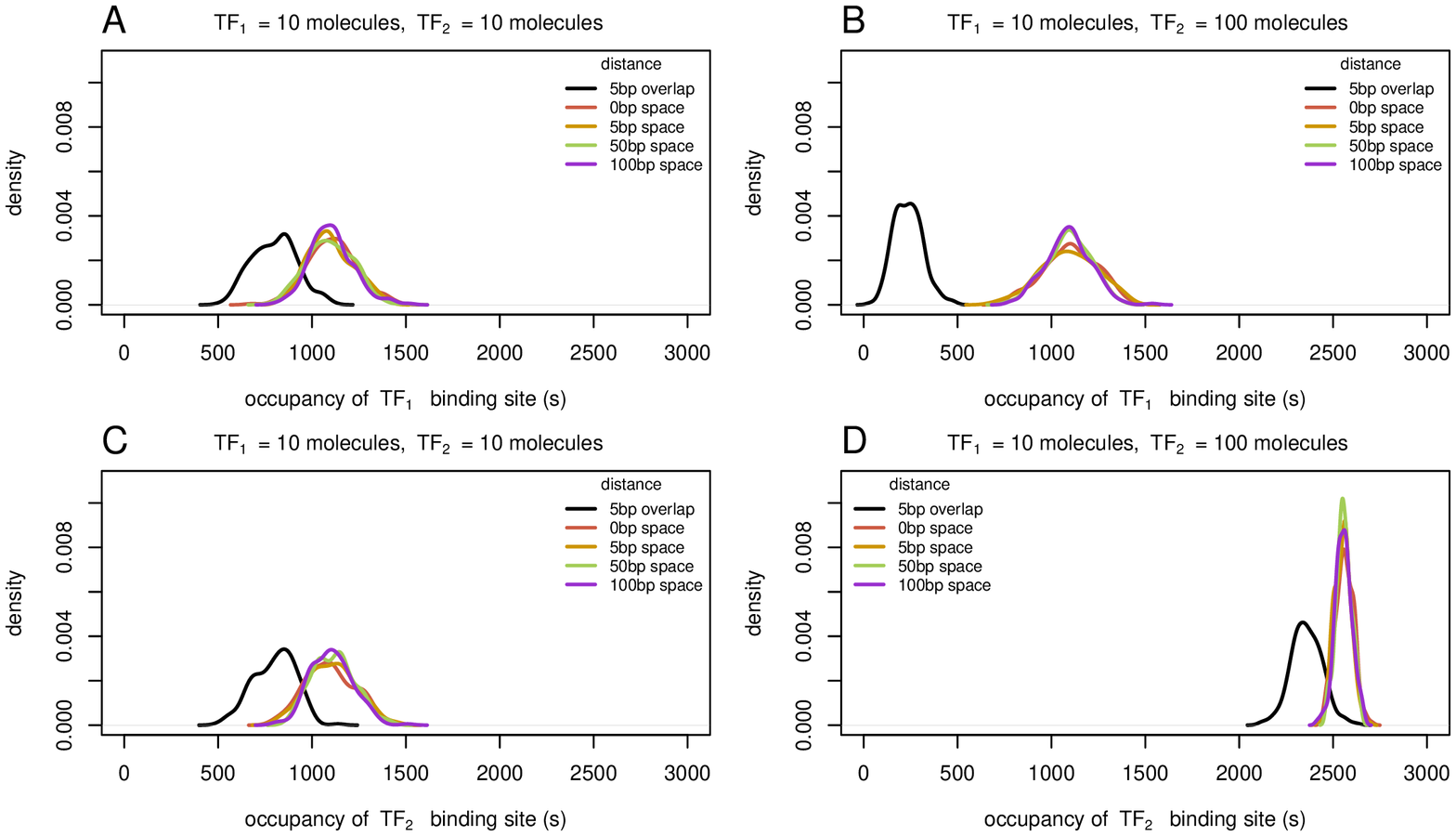}
\caption{\emph{Occupancy of switches and barriers in GRiP}. Here we simulated the switch and barrier scenarios with various distances between the binding sites (for switch, $5\ bp$ overlap and, for barriers, $\in\{0, 5,50,100\}\ bp$ space between the two binding sites). We measured the occupancy (the total amount of time a binding site is bound during a period of time) across an \emph{E. coli} cell cycle ($3000$ seconds) when \onel\ and \threel\ $TF_1=TF_2=10\ molecules$ and \twol\ and \fourl\ $TF_1=10\ molecules$ and $TF_2=100\ molecules$.   Binding sites in the switch configuration have significantly less occupancy than binding sites that are far apart and higher variability in occupancy. On the other hand, barriers do not seem to have a significant effect on occupancy.  \onel\ and \twol\ the occupancy for $TF_1$. \threel\ and \fourl\ the occupancy for $TF_2$.\label{fig:barrieroccupancy}}
\end{figure}

\begin{figure}
\centering
\includegraphics[width=\textwidth]{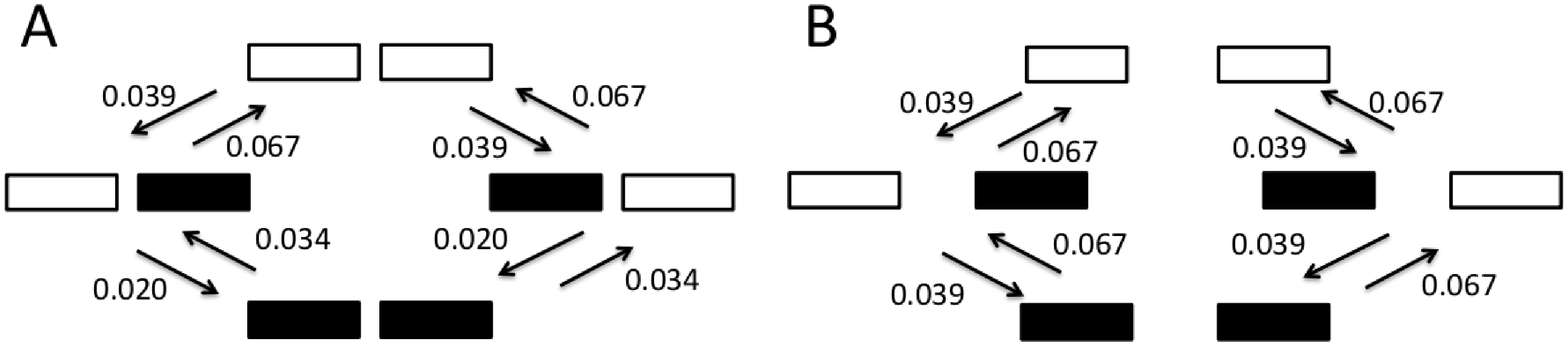}
\caption{\emph{fastGRiP transition diagrams for barriers and far apart binding sites}.  One of the reasons why overall TF occupancy is not significantly affected by the \emph{barrier effect} is that the both the rate of binding and unbinding events is reduced in the barrier case.  Here we show the Markov Chains used in fastGRiP that describe a barrier \onel\ and two far apart binding sites \twol.  In fastGRiP, the states represent specific configurations (white boxes are unbound sites and black boxes are bound sites) and transitions represent specific binding or unbinding reactions.  Along the arrows that represent the state transitions, we show the propensities of binding and unbinding events, as calculated by fastGRiP for \onel\ barriers  and \twol\ far apart binding sites . The expected reaction time is inversely proportional to the reaction propensity.  We see that the barriers case has lower or equal reaction propensities for all TF binding and TF unbinding reactions compared to the case of far away binding sites.  Therefore, there will be less frequent binding and unbinding events if the TF binding sites are closely spaced.  Even though we saw in  that the overall occupancy (the total time a TF is bound across a cell cycle) is not significantly affected by barriers  (Figure \ref{fig:barrieroccupancy}), we see that barriers influence the TF binding kinetics. \label{fig:barriertransitions}}
\end{figure}

\clearpage
\begin{figure}
\centering
\includegraphics[angle=270,width=\textwidth]{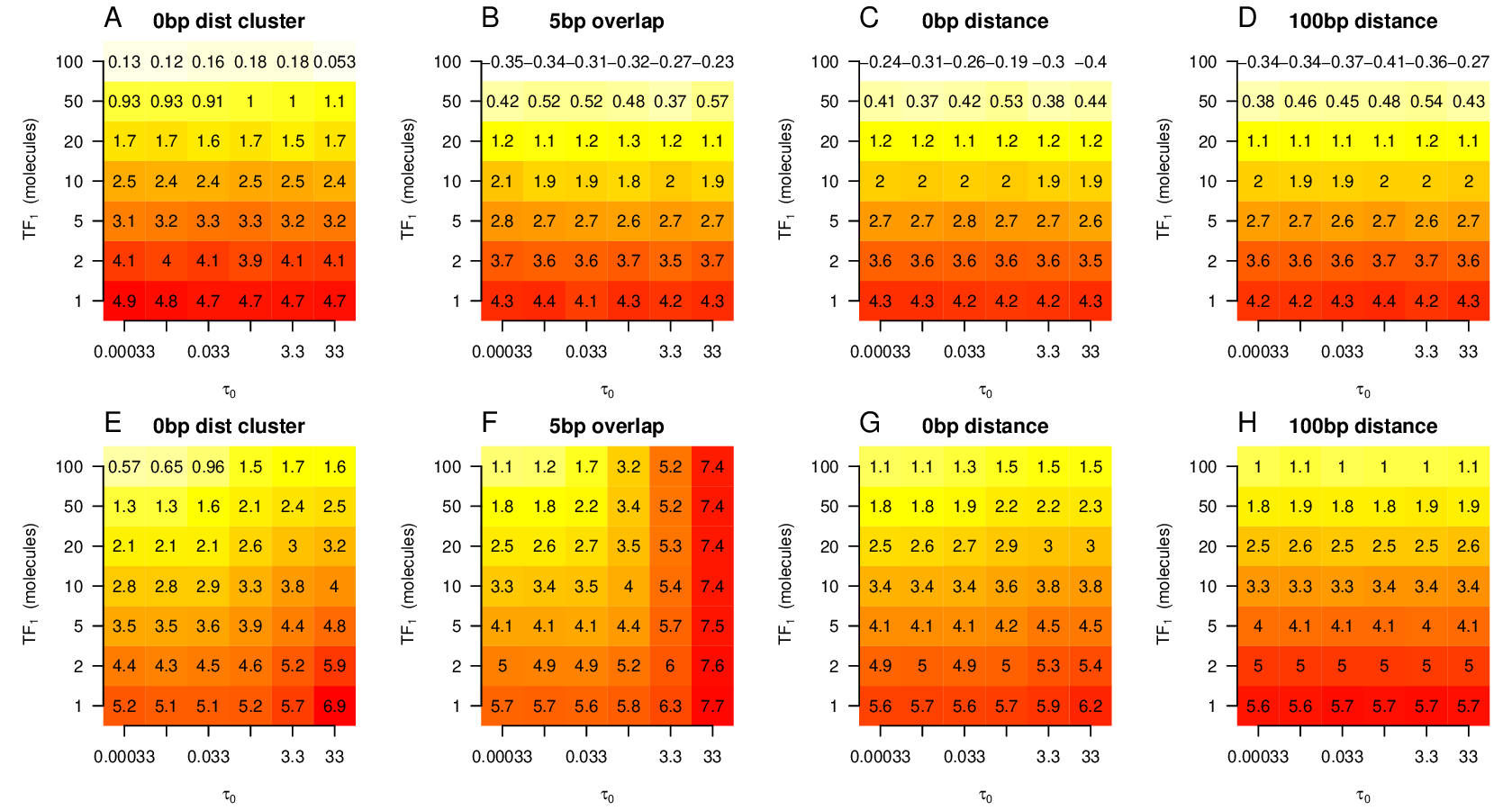}
\caption{\emph{TF arrival time in fastGRiP for symmetric system}. We change the abundance and binding affinity of \emph{both} TFs in the system symmetrically.  In other words, we simulate a pair of co-localised binding sites, changing the binding affinities and abundances of both TFs in the system.  We observe the arrival time (in ln(seconds)) for the first of the two TFs, $ln(min(search\ time\ TF_1, search\ time\ TF_2))$, in ($A-D$) and the last of the two TFs, $ln(max(search\ time\ TF_1, search\ time\ TF_2))$, in ($E-H$).  We ran this experiment on \onel\ and \fivel\ clusters ($0\ bp$ distance between sites);  \twol\ and \sixl\ switches ($5\ bp$ overlap between sites); \threel\ and \sevenl\ barriers ($0\ bp$ distance between sites) and \fourl\ and \eightl\ far away binding sites ($100\ bp$ distance between sites).  As before, the values represent the averages across $400$ simulations.  We see that TF abundance influences the time of binding of the first arriving TF, but binding affinity does not affect the first TF's arrival time.  The first arriving TF in a cluster is slower than the switch, barrier, or far away cases, because a neighbouring binding site in a cluster can act like a trap (because the TF will bind to the first TF binding site it bumps into).  As shown in \sixl\ and \sevenl, in the switch case, and to a lesser extent in the barrier case, increasing binding affinity slows the binding of the latest arriving TF.  Most importantly, in \fivel\ we can see the dual nature of cluster: at high levels of binding affinity and low concentrations, the binding of the second TF is slowed.  On the other hand, at low binding affinities and high concentrations, the second TF will bind faster in the case of a cluster than it would in the case of the two binding sites being far apart, because the TF can slide back and forth between the two binding sites.} \label{fig:clusterbuildingblock_heatmap}
\end{figure}

\clearpage
\begin{figure}
\centering
\includegraphics[width=\textwidth]{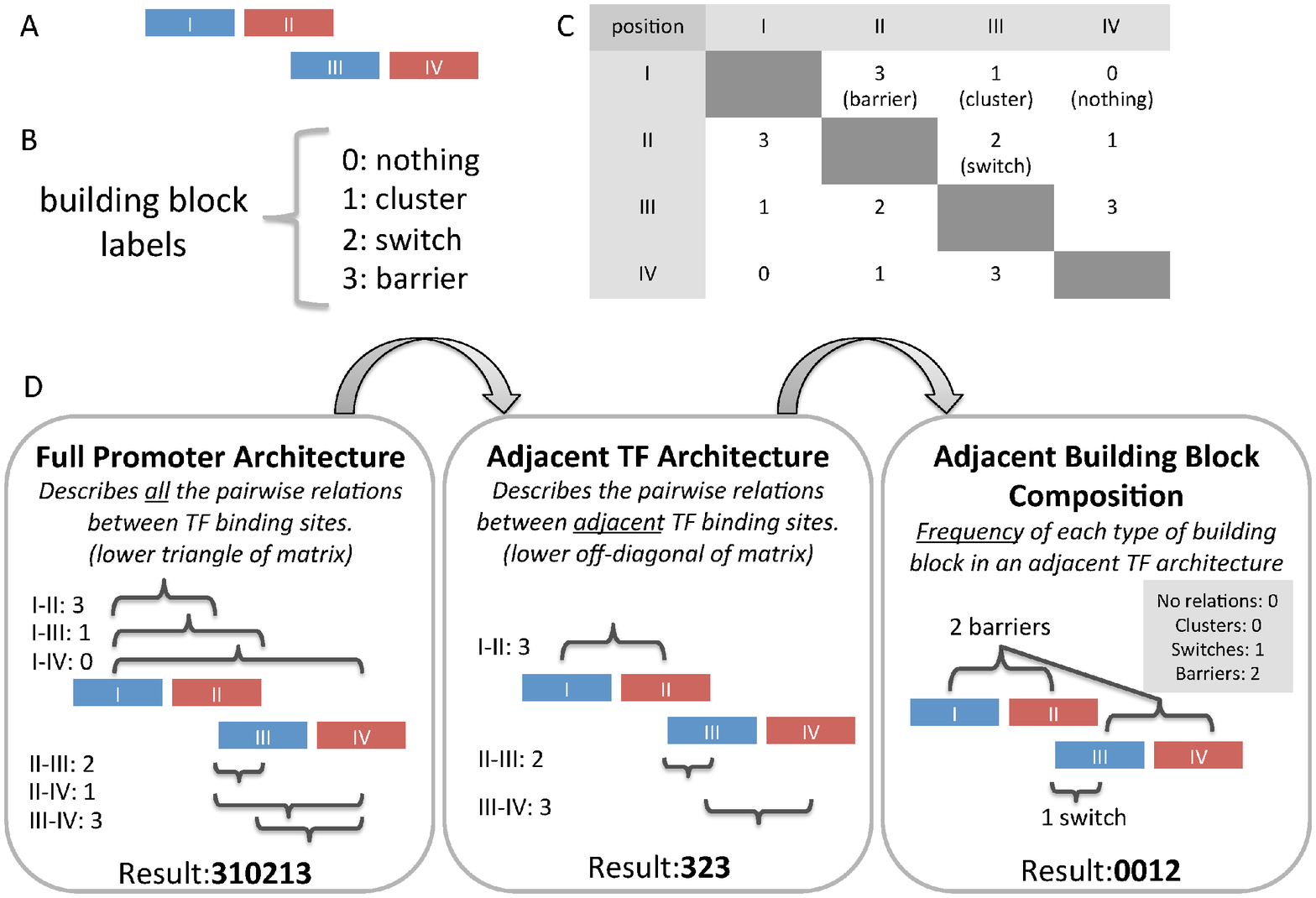}
\caption{\emph{Promoter architecture notation}.  It is extremely useful to classify promoter organisations by their composition of switches, barriers, and clusters in an unambiguous and enumerable way. Here we present a versatile scheme for enumerating promoter architectures. (A) shows an example promoter architecture, which we will use as an example to demonstrate the algorithm we use to categorise them, comprising binding sites I to IV. (B) shows our mapping of numerals to building blocks, which we can use to label a particular pairwise interaction. (C) shows all of the pairwise interactions between TFs for our example case. For instance, as we see that binding site I and binding site II form a barrier, we put the number 3 in the corresponding location in the table. In (D), we describe how we can transform this table into numerical labels for the promoter architecture. First, we define the \emph{full promoter architecture label} to describe all pairwise interactions between TFs and their binding sites. We show the specific order of comparisons we make in order to have a consistent labeling scheme. Note that this corresponds to the lower triangular portion of the table in (B), read column by column. Next, we define the \emph{adjacent TF architecture label} as the description of all pairwise interactions between TFs that are immediately adjacent to one another. This corresponds to the upper or lower off-diagonal of the table. In this example, we see clusters are part of the full promoter architecture, but not part of the adjacent TF architecture, because binding sites I and III nor binding sites II and IV are immediately adjacent to one another. These relationships between adjacent TFs often contribute the most to the behavior of the promoter architecture. Finally, we refer to the counts of each type of building block as the \emph{adjacent building block composition label}. This label represents the frequency of observing switches, barriers, and clusters among adjacent pairs of binding sites. Our enumeration strategy is particularly useful because: it defines a way of easily label promoter architectures by its building blocks and also defines a hierarchy of promoter organisations (from promoter architecture to adjacent binding sites architecture to building block composition). This helps group promoters based on their architectures and look at the distribution of architectures across the genome more effectively. (Note that if the promoter architecture was drawn in the opposite orientation, then the table could change, so to keep consistency, we always take the label that has the highest value.)  \label{fig:notation}}
\end{figure}

\clearpage
\begin{figure}
\centering
\includegraphics[width=\textwidth]{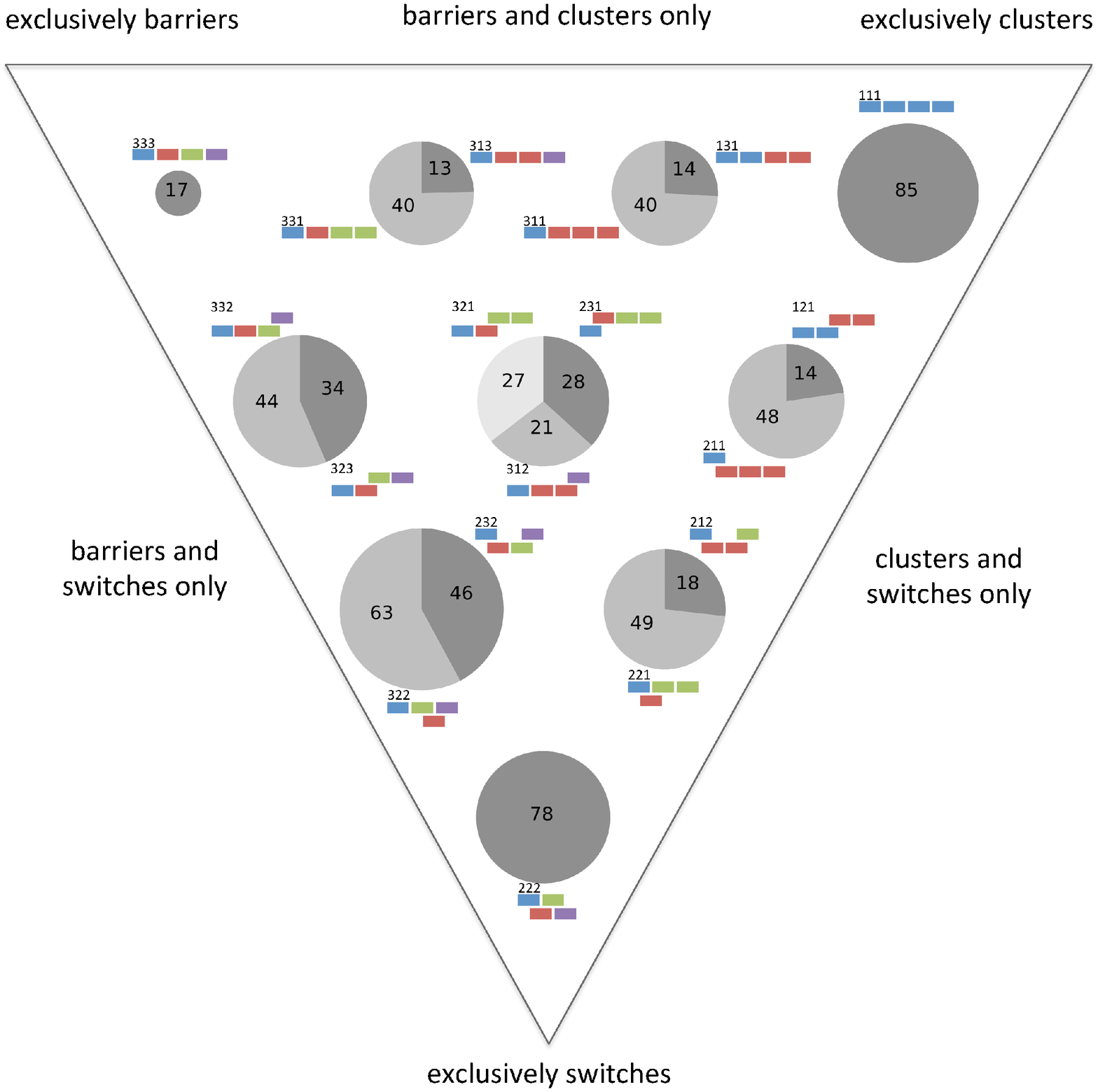}
\caption{\emph{Distribution of sets of 4 binding sites in the \emph{E. coli} genome}.  We show the distribution of all sets of four binding sites $<100\ bp$ apart.  Each circle represents a different building block composition, and the proximity of each circle to each corner corresponds with how many barriers, switches, and clusters are present.  For instance, the circles in the corners represent architectures that consist of only one type of building block (such as, only barriers).  The circle in the center represents architectures with one of each type of building block (one barrier, one cluster and one switch).  The size of each circle represents the number of times we observe that building block composition in the \ecoli\ genome.   Within each circle is a pie chart that shows the relative frequencies of each adjacent TF architecture.   The numbers inside the pie chart correspond to the number of times that adjacent TF architecture was found in the \ecoli\ genome.  Each of the architectures are also drawn, and labelled with their associated adjacent TF architecture label.  \label{fig:tripletdist}}
\end{figure}

\clearpage
\begin{figure}
\centering
\includegraphics[width=\textwidth]{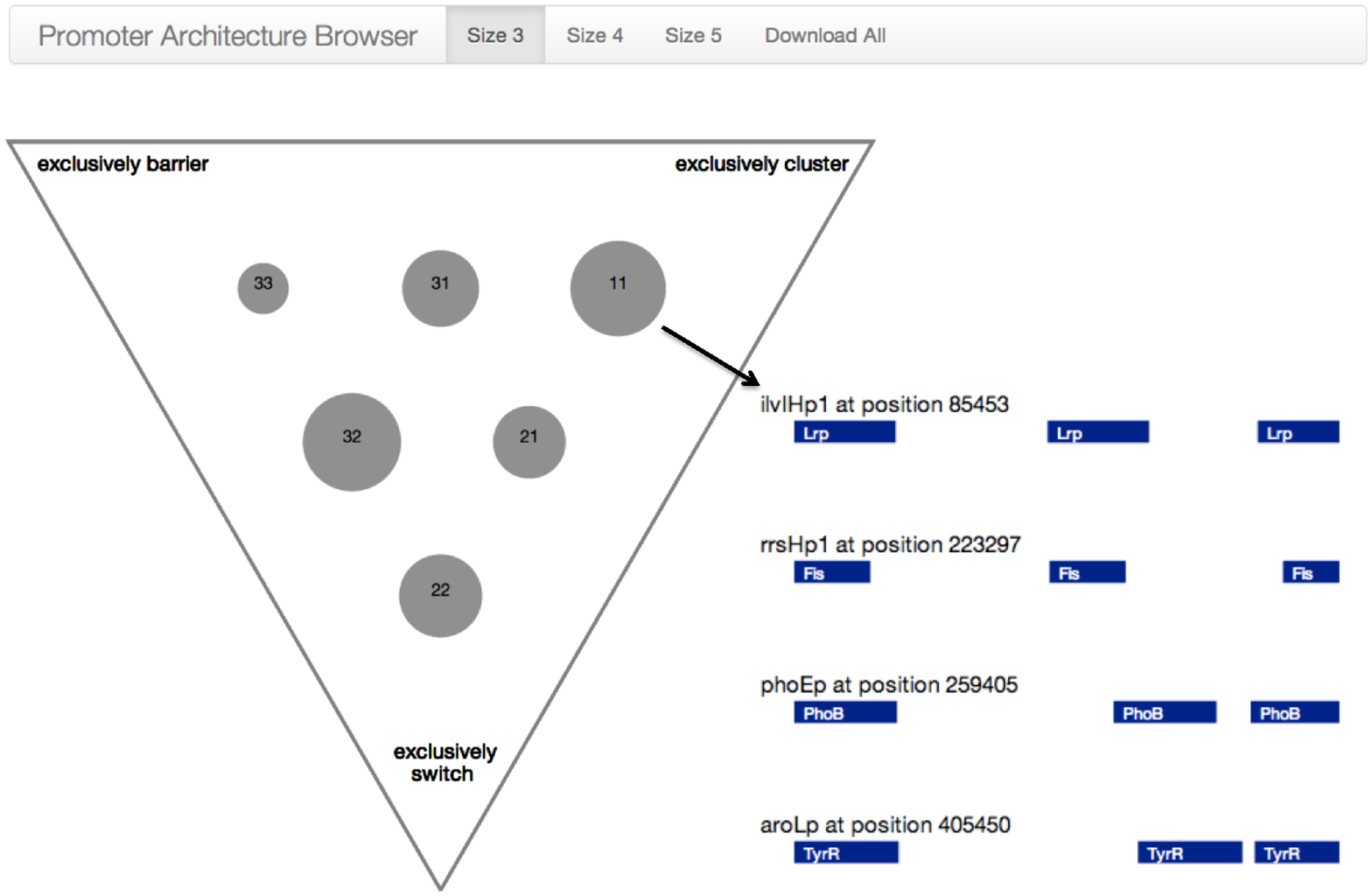}
\caption{\emph{Webtool for browsing binding site architectures in the \emph{E. coli} genome}.  We have classified promoter architectures of every set of 3, 4, and 5 TF binding sites that are less than 100 bp apart from one another. To easily browse the \ecoli\ architectures, we have created a website (http://logic.sysbiol.cam.ac.uk/fgrip/db) that allows the user to click of each part of the pie chart and scroll through the corresponding \ecoli\ promoters.
\label{fig:fivemerdist}}
\end{figure}

\clearpage

\begin{figure}
\centering
\includegraphics[angle=-90, width=\textwidth]{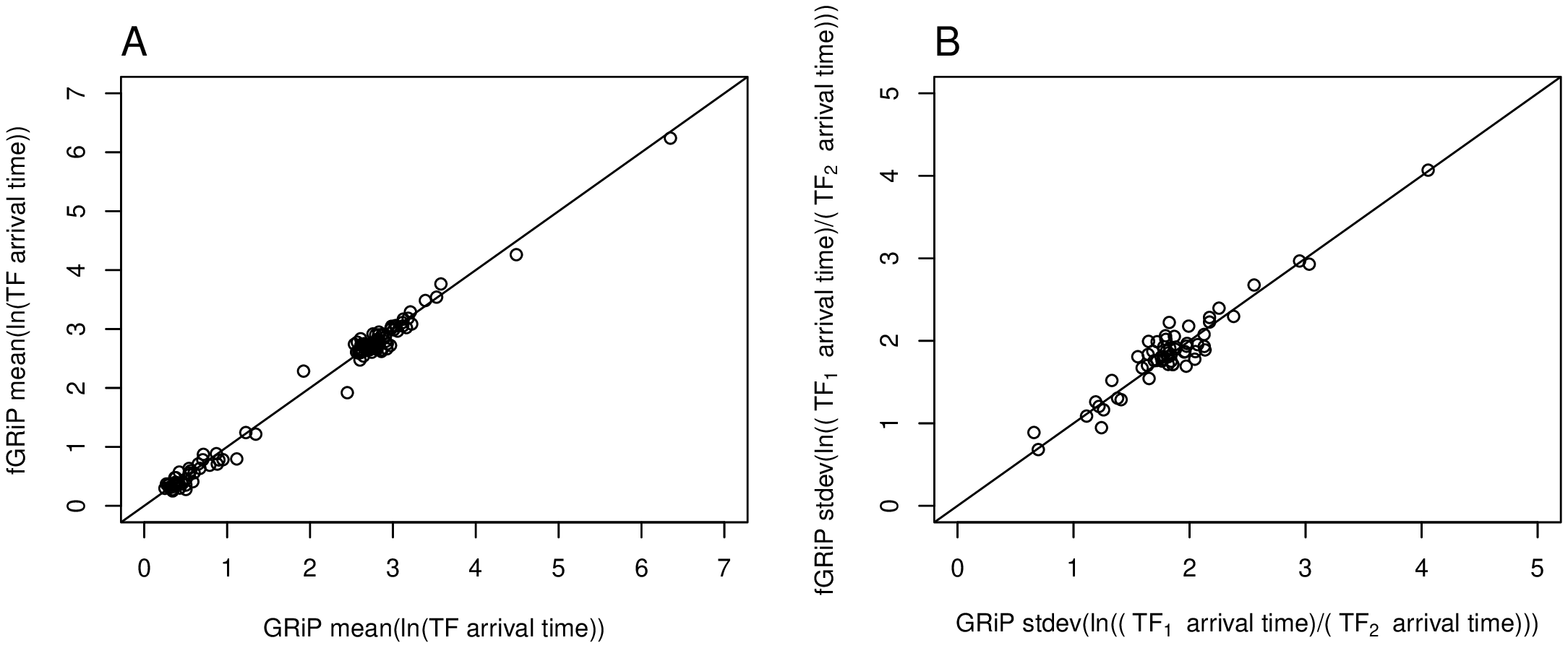}
\caption{\emph{Comparison between GRiP and fastGRiP}. Here we compare the mean of the log of the arrival times (A)  and the standard deviation of the log of the arrival ratios (B) between GRiP, the fully stochastic simulation, and fastGRiP, the semi-analytic approximation.  Each point represents one building block experiment (a switch, barrier, or cluster) under a specific parameter set: a TF abundance of 10 or 100, a binding affinity $t_0$ value of 0.033, 0.33, or 3.3, and a distance between binding sites of -5 (overlap), 0, 5, 50, or 100. (A) shows that the values for the arrival times are consistent between the simulations, with $R=0.995$. (B) shoes that the relative arrival times are consistent between the simulations, with $R=0.954$. \label{fig:comparisonToGRiP}}
\end{figure}

\begin{figure}
\centering
\includegraphics[width=\textwidth]{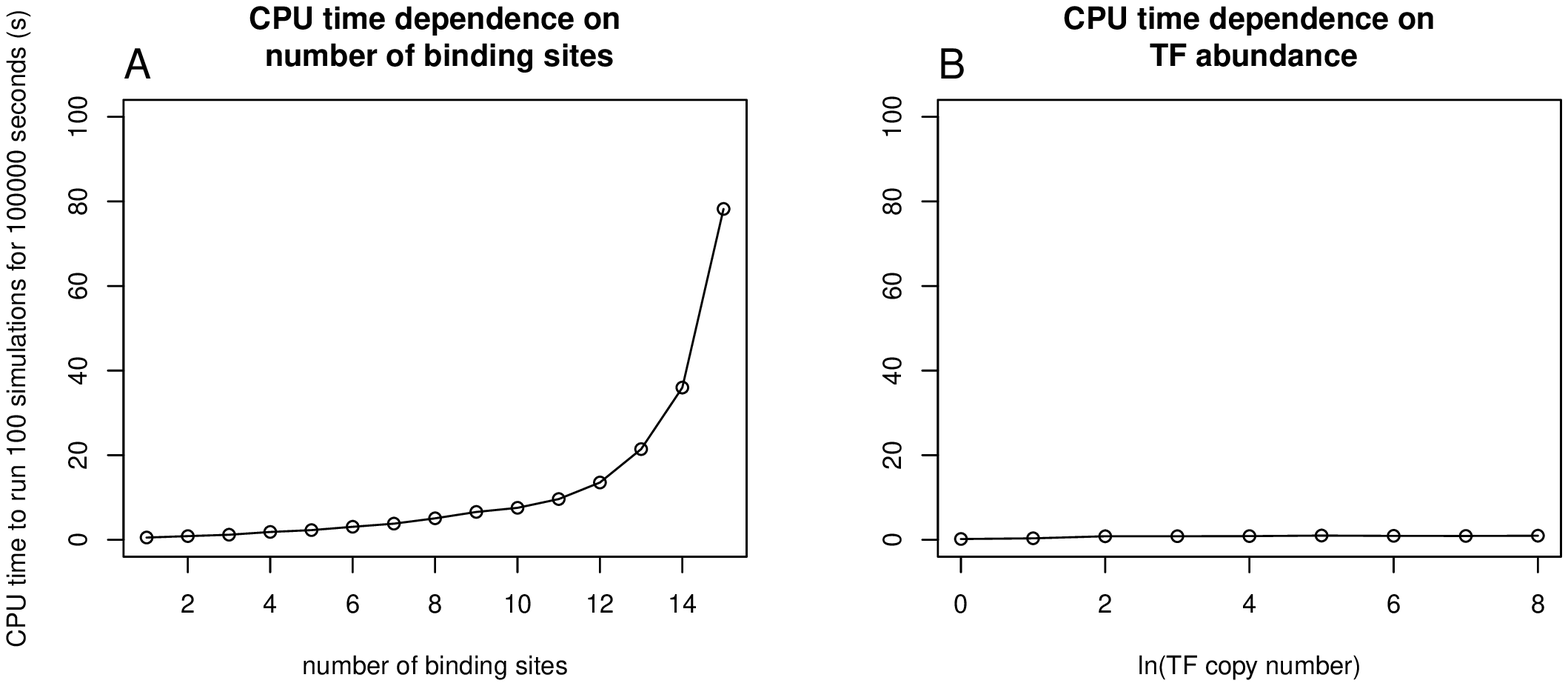}
\caption{\emph{Time required to simulate $10^{5}\ s$ with fastGRiP}. The simulations were performed on a MacBook Air 1.8 GHz Intel i5 CPU with 4GB memory running Mac OSX 10.7.  \onel\ The dependence of the the CPU time on the  number of binding sites, in the case of $1000$ molecules. fastGRiP becomes significantly slow for high number of binding sites. However, for usual number of binding sites in bacterial promoters, fastGRiP is fast. \twol\ The dependence of CPU time on the number of molecules, in the case of $2$ binding sites. Practically,  fastGRiP simulation time is not affected by the number of molecules. This is opposite to GRiP where there is linear dependence with the length of the DNA (and, thus, with the number of binding sites) and exponential dependence on the number of molecules.  \label{fig:runtime}}
\end{figure}

\clearpage

\begin{figure}
\centering
\includegraphics[angle=-90, width=\textwidth]{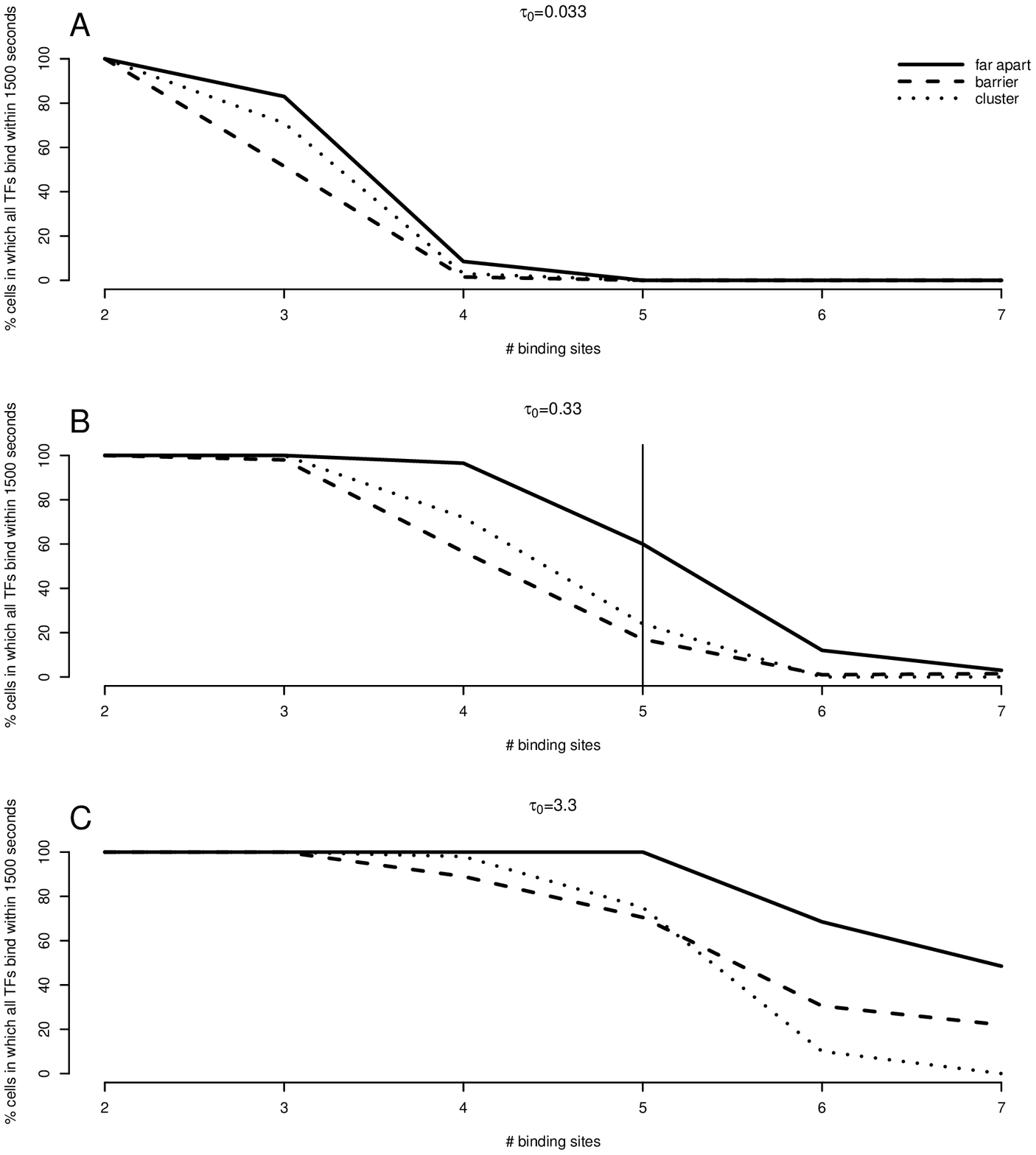}
\caption{\emph{Influence of promoter architecture on formation of AND configuration.} Here we show the proportion of cells (simulations) in which the AND configuration (all TFs bound at once) is reached before half a cell cycle (1500 seconds), dependent on the number of binding sites and the organisation of the promoter (TFs far apart and therefore unaffected by facilitated diffusion-- solid line, TFs 0bp apart in a barrier organisation-- dashed line and TFs 0bp apart in a cluster organisation-- dotted line).  Each graph shows the results at a different level of binding affinity ($\tau_0=0.033$ in (A), $\tau_0=0.33$ in (B) and $\tau_0=3.3$ in (C)), but a constant TF abundance of 10 molecules.  As expected, in all cases the higher the complexity of the promoter, the less likely it is for all TFs to find their binding sites within half a cell cycle; however, when the TFs are far apart from one another, a higher proportion of cells would reach the AND configuration within half a cell cycle than in the case of barriers or clusters.  For example, as indicated by the horizintal line in (B), in the case of a $\tau_0$ of 0.33 and 5 binding sites, the number of cells that reach the AND configuration within half a cell cycle is reduced by more than half between the far away case and the barrier/cluster cases (from $60\%$ to $24\%$).  Notice that the higher the binding affinity between the TFs and the DNA, the larger the effect of closely packing TF binding sites on the formation of the AND configuration.  At lower binding affinities, the relative effect is smaller, but one begins to see an effect when there are fewer adjacent binding sites.
\label{fig:biologicallyrelevant}}
\end{figure}

\begin{figure}
\centering
\includegraphics[width=\textwidth]{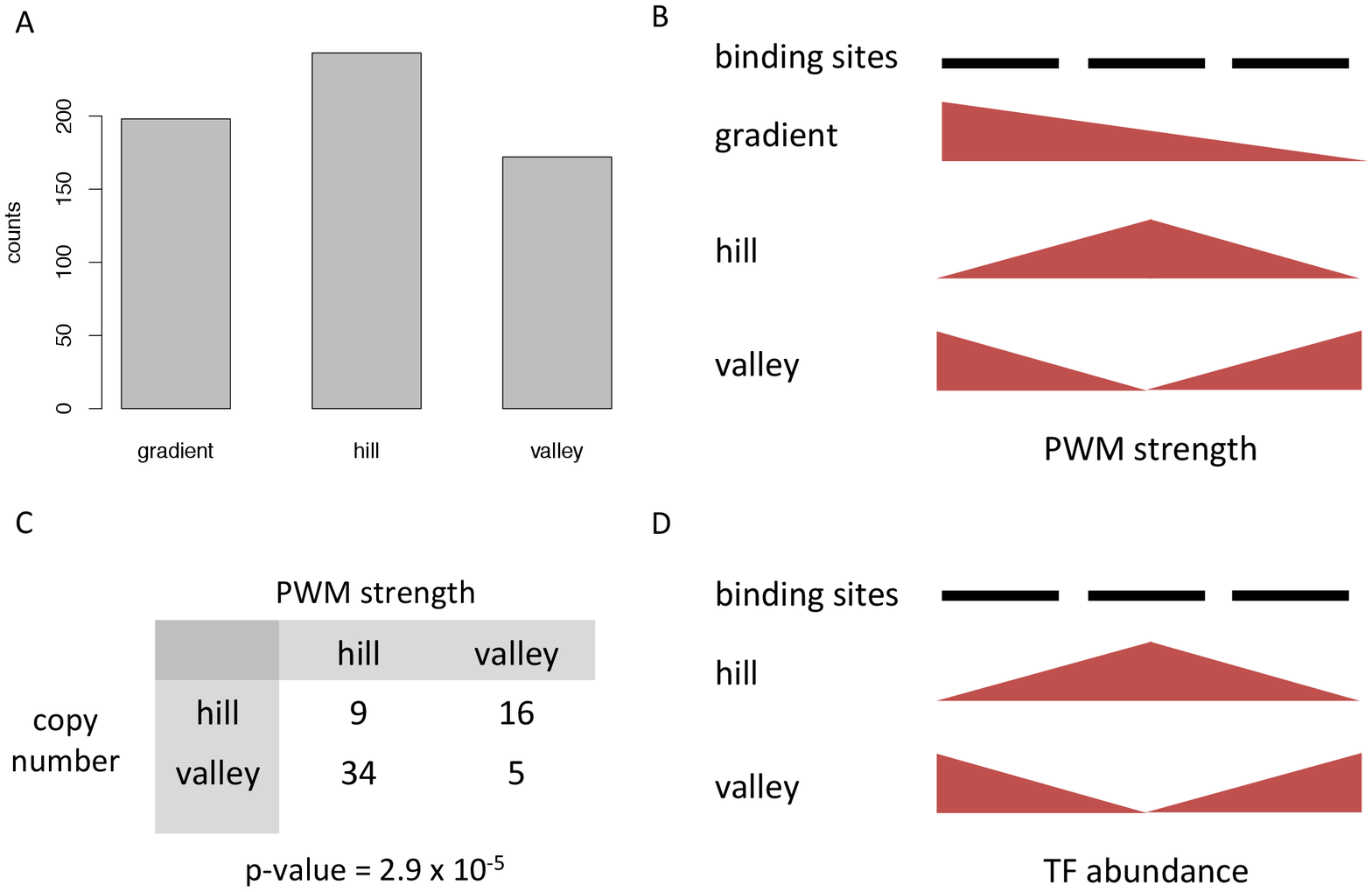}
\caption{\emph{Hill and valleys in E. coli genome}. When three binding sites are closely spaced, the relative binding affinities and abundances of the outer binding sites compared to the central binding site influences the rate at which all three binding sites can be bound at once (\emph{AND configuration}).  We define a \emph{hill} to be a configuration in which the central binding site has a higher binding affinity \twol\ or abundance \fourl\ and a \emph{valley} to be a configuration in which the central binding site has the lowest binding affinity \twol\ or abundance \fourl. Other configurations are classified as \emph{gradients}. We categorized binding sites triplets in \emph{E. coli strain K-12}  that were $<50 bp$ apart from one another as hills, valleys, or gradients in terms of PWM scores \onel\ and abundances \threel. \onel\ The distribution of hills, valleys, and gradients in terms of PWM scores is significantly different from the expected distribution (p-value$=0.0018$, chi-squared test), with an enrichment for hills.  When we calculated hills and valleys in terms of abundances of TFs \threel, we only considered triplets that had two or more TFs that had high enough abundances such that their concentrations could be measured by APEX \cite{Lu2007}. Hills and valleys in terms of binding affinity and concentration appear to be anti-correlated and the results vary significantly from the null distribution (p-value$=2.9\times 10^{-5}$, Fisher exact test). Triplets that are hills in terms of binding affinity and valleys in terms of concentration are best suited for having multiple TFs bound at once, and this seems to be the case within the \ecoli\ genome. \label{fig:hillvalleybio}}
\end{figure}

\clearpage

\begin{figure}
\centering
\includegraphics[width=\textwidth]{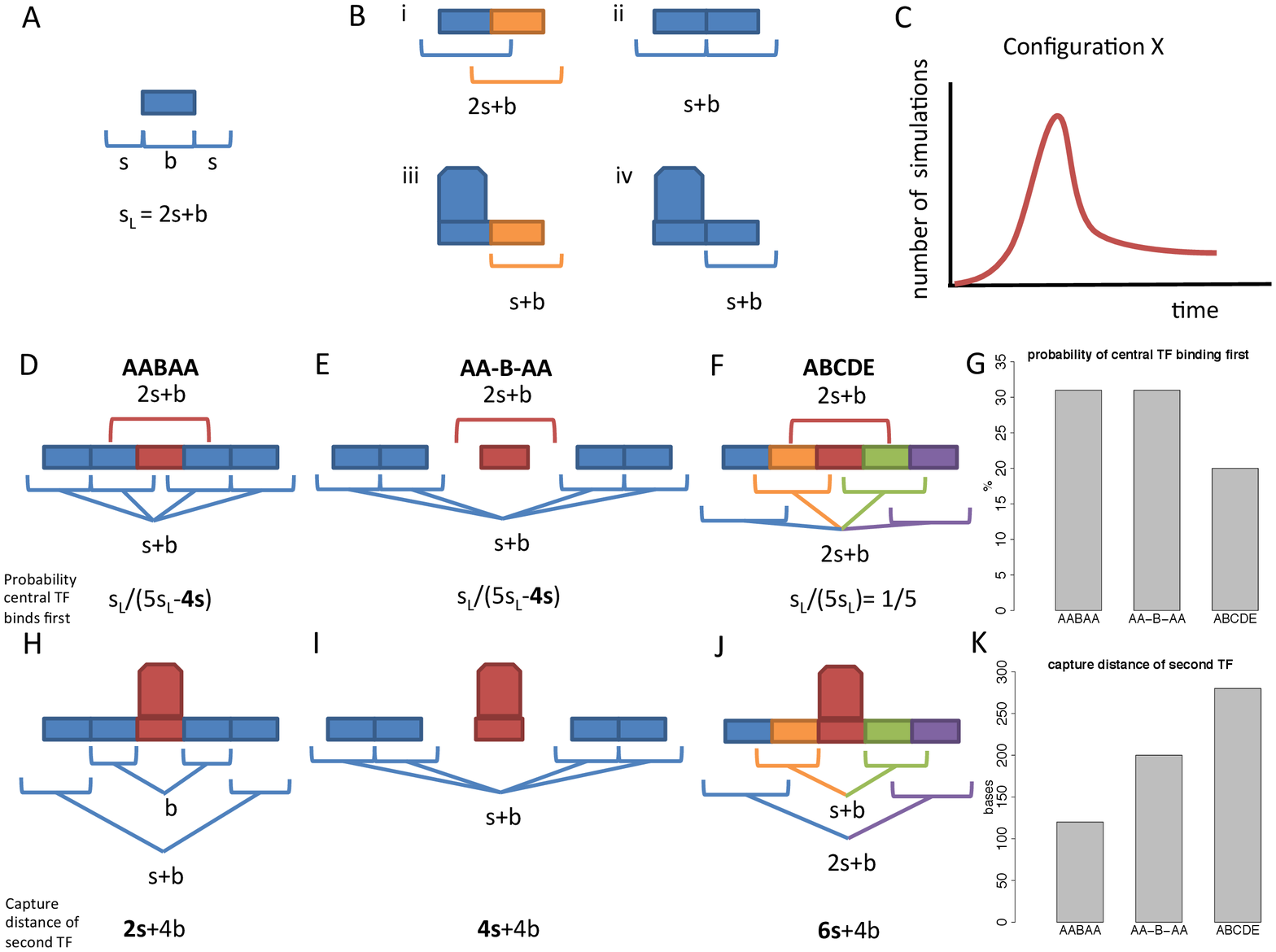}
\caption{\emph{Mechanism underlying impulse behavior.} In (A) the sliding length ($s_L$) is defined as the length of the DNA segment where a TF will almost certainly find its binding site by 1D random walk.  For the purpose of this figure, we can describe $s_L$ as the sum of the length of the binding site $b$ and a region on either side of the binding site, $s$ bases on each side.  (B) Here we illustrate how the sliding length varies in the case of barriers (i, iii) and clusters (ii, iv), and when the adjacent site is unbound (i, ii) or bound (iii, iv).  Subfigures (iii, iv) re-iterate the barrier effect, while subfigure (ii) illustrates that a TF will bind to the first site it reaches, and therefore \emph{unbound} binding sites in a cluster act like barriers.  (C)  A configuration is defined as the combination of TFs that are bound to the DNA at a given time point (such as the case of $TF_B$ being bound, but no $TF_A$ being bound in an AABAA architecture). We define an impulse behavior as a short period of time in which there is a higher probability of a certain configuration occurring than observed at equilibrium   Note that two of the phenomenon that influence this include 1) how often the central TF binds first, which we will describe further for the AABAA, AA-B-AA, and ABCDE cases in figures (D-G) and 2) the time delay before additional TFs bind, which we explore further in figures (H-K).  In (D-F) we illustrate the sliding widths for each binding site in the case of empty DNA.  Figures (D-F) show how there is a relative higher probability of the central TF becoming bound in the AABAA and AA-B-AA cases (D, E) compared to the ABCDE case (F).  Using $b=10$ and $s=40$ (biologically relevant parameters), we show how the probability of the central TF binding first differs between the three cases.  In (H-K), we illustrate the total sliding widths for each binding site when only the central TF is bound.  The AABAA configuration has the lowest total sliding width (H) compared to the AA-B-AA and ABCDE (I, J) cases, which implies that the AABAA configuration would result in the longest time delay before a second TF becomes bound.  Figure (K) compares total capture distance (region in which a TF will likely bind to one of the sites if it binds there) in each of the three cases with the same parameters as in (G). This is one way of interpreting the underlying mechanism that results in AABAA having the largest magnitude of an impulse, while AA-B-AA has a lower magnitude of an impulse and ABCDE barely has an impulse at all.  
\label{fig:mechanism}}
\end{figure}

\begin{figure}
\centering
\includegraphics[width=\textwidth]{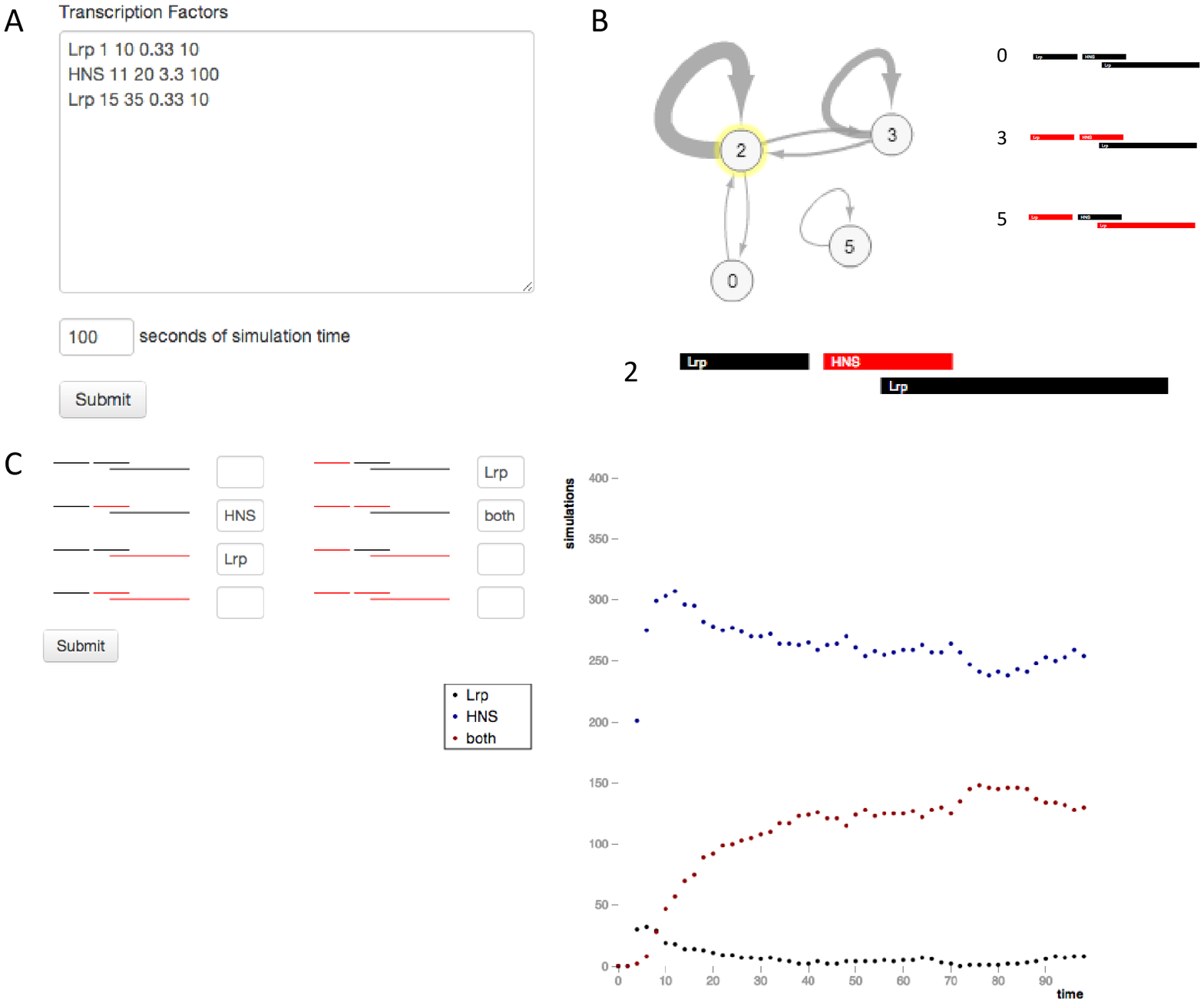}
\caption{\emph{fastGRiP website user interface.} Here we show screen captures from an exemplary promoter (http://logic.sysbiol.cam.ac.uk/fgrip/).  In (A) we see the input screen, filled with demo data.  The TF names, start and end positions of the binding sites, the $\tau_0$ binding affinity parameter, and TF abundance must be specified, along with the number of seconds of the \ecoli\ life cycle for the simulation to run.  In (B), we see a network output for fastGRiP, where each node represents a specific promoter configuration (for instance, configuration 3, has the first two TFs bound).  The thickness of the arrows indicates the frequency of a particular configuration-to-configuration transition across 5 seconds.  For instance, if the cell is in configuration 2 (HNS only bound), then it will likely continue to be in configuration 2 after 5 seconds.  This network helps illustrate the dynamics of how promoter configurations change over time.  When one clicks on a node, it gets highlighted (as shown for config 2), and the configuration map is illustrated.  In (C), we show another graphical output of fastGRiP.  All the possible configurations are illustrated to the left, with a place in which the user can specify labels for the configurations of interest.  After clicking the submit button, a scatterplot will graph the frequency of those configurations over time.  If two configs have the same label, the scatterplot will display the sum of the frequency of their configurations.  
\label{fig:fastGRiP}}
\end{figure}

\clearpage
\begin{table}

\centering
  \begin{tabular}{|>{\centering}m{2.5cm}| m{12.5cm}| }
        \hline
        \textbf{parameter} &  \textbf{description} \\ \hline
	$M$ & the length of the DNA; $M\approx4.6\ Mbp$ \cite{riley_2006}. \\ \hline
	$f$ & the proportion of time a molecules spends on the DNA; $f\approx 0.9$ \cite{elf_2007}.  \\ \hline	
	$t_R$ & residence time, the amount of time a TF performs a one dimensional random walk on the DNA, before it unbinds; $t_R=5\ ms$ \cite{elf_2007}. \\ \hline
	$k^{\textrm{dissoc}}$ & the dissociation rate constant between a TF molecule and the DNA; $k^{\textrm{dissoc}}_x=1/t_R=200$ \cite{zabet_2012_model}. \\ \hline
	$k^{\textrm{assoc}}_{x}$ & the association rate constant for a TF to the DNA; for the full system, $k^{\textrm{assoc}}_{x}=2400\ s^{-1}$ \cite{zabet_2012_model}, while for smaller system we used equation 10 in \cite{zabet_2012_subsystem}.\\ \hline
	$s_l$ & sliding length, the average number of positions on the DNA scanned during a one dimensional random walk; $s_l=90\ bp$ \cite{elf_2007}. \\ \hline
	$P_{\textrm{unbind}}$ & the probability to unbind from the DNA; $P_{\textrm{unbind}}=0.00147$ \cite{zabet_2012_model}. \\ \hline
	$P_{\textrm{left}}$, $P_{\textrm{right}}$ & the probability to slide left ($P_{\textrm{left}}$) or right ($P_{\textrm{right}}$) on the DNA; $P_{\textrm{left}}=P_{\textrm{right}}=0.4994$ \cite{zabet_2012_model}. \\ \hline
	$P_{\textrm{jump}}$ &  the probability that the molecule releases into the cytoplasm during an unbinding event, while $(1-P_{\textrm{jump}})$ represents the probability that a molecule will rebind fast after a dissociation (hop); $P_{\textrm{jump}}=0.1675$ \cite{wunderlich_2008}. \\ \hline
	$\sigma^{2}_{\textrm{hop}}$ &  the variance of the hop distance, which is Gaussian distributed around unbinding position;$\sigma^{2}_{\textrm{hop}}=1\ bp$ \cite{wunderlich_2008}. \\ \hline
	$d_{\textrm{jump}}$ &  the distance over which a hop becomes a jump; $d_{\textrm{jump}}=100\ bp$  \cite{wunderlich_2008}.  \\ \hline
	$\tau_{x}^{0}$ & the average waiting time for species $x$ when bound specifically; here we used the value of $\tau_{x}^{0}=0.33$ which was derived in \cite{zabet_2012_model} for non-cognate TFs. \\ \hline
	$\zeta$ & the pseudo-count term for the PWM; $\zeta=1$ \cite{berg_1987}. \\ \hline
	$TF^{\textrm{size}}_{x}$ & the number of base pairs covered by a bound TF molecule; $TF_{\textrm{nc}} \geq 20\ bp$ \cite{stormo_1998}.\\ \hline
	$TF^{\textrm{motif}}_{x}$, $TF^{\textrm{left}}_{x}$ and $TF^{\textrm{right}}_{x}$ & the number of base pairs covered  by the DNA binding domain ($TF^{\textrm{motif}}_{x}$), to the left ($TF^{\textrm{left}}_{x}$) of the DNA binding domain and to the right  ($TF^{\textrm{right}}_{x}$) of the DNA binding domain. When we provide the affinity landscape  to GRiP (thus, not specifying $TF^{\textrm{size}}_{x}$), we use the $TF^{\textrm{left}}_{x}$ and $TF^{\textrm{right}}_{x}=0$ to denote that a bound molecule covers $TF^{\textrm{left}}_{x}+TF^{\textrm{right}}_{x}=0$ base pairs on the DNA.\\ \hline	
	$K_B$ and $T$ & Boltzmann constant ($K_B$) and temperature ($T$). \\ \hline
  \end{tabular}
\caption{\emph{Nomenclature}. This is a subset of the parameters associated with the facilitated diffusion, which were listed in \cite{zabet_2012_model}.}\label{tab:paramsDescription}
\end{table}

\clearpage
\begin{table}
\centering
\begin{tabular}{| l | r | r |r |}
\hline
\textbf{parameter} &  \textbf{$TF_1$} & \textbf{$TF_2$} & \textbf{notation}\\ \hline
copy number &  $10$ & $10$ & $TF_{x}$\\ \hline
motif sequence &  \verb+C+ & \verb+G+ & \\ \hline
energetic penalty for mismatch &  $16K_BT$ & $16K_BT$ & $\varepsilon^{*}_{x}$\\ \hline
nucleotides covered on left &  $10\ bp$ & $10\ bp$ & $TF^{\textrm{left}}_{x}$\\ \hline
nucleotides covered on right &  $10\ bp$ & $10\ bp$ & $TF^{\textrm{right}}_{x}$\\ \hline
association rate to the DNA &  $0.86\ s^{-1}$ & $0.86\ s^{-1}$ & $k^{\textrm{assoc}}_{x}$\\ \hline
unbinding probability &  $1.47E-3$ & $1.47E-3$ & $P^{\textrm{unbind}}_{x}$\\ \hline
probability to slide left &  $0.4992629$ & $0.4992629$ & $P^{\textrm{left}}_{x}$\\ \hline
probability to slide right &  $0.4992629$ & $0.4992629$ & $P^{\textrm{right}}_{x}$\\ \hline
probability to dissociate completely when unbinding &  $0.1675$ & $0.1675$ & $P^{\textrm{jump}}_{x}$\\ \hline
time bound at the target site &  $0.33\ s$ & $0.33\ s$ & $\tau^{0}_{x}$\\ \hline
the size of a step to left &  $1\ bp$ & $1\ bp$ & \\ \hline
the size of a step to right &  $1\ bp$ & $1\ bp$ & \\ \hline
variance of repositioning distance after a hop &  $1\ bp$ & $1\ bp$ & $\sigma^{2}_{\textrm{hop}}$\\ \hline
the distance over which a hop becomes a jump  &  $100\ bp$ & $100\ bp$ & $d_{\textrm{jump}}$\\ 
\hline
\end{tabular}
\caption{\emph{TF species default parameters}. We consider the case of two TF species $TF_1$ and $TF_2$ and a uniform affinity landscape. This is implemented by assuming that the DNA is a string form only of adenine and at the target site (usually position in the middle of the DNA) we consider that the DNA contains the 1-bp recognition motifs for the two TFs, namely: cytosine for $TF_1$ and guanine for $TF_2$. We considered a $20\ Kbp$ DNA segment which is smaller compare to the \emph{E.coli}-K12 genome (which is $4.6\ Mbp$) and, thus, we applied the association rate model to adapt the association rate of the small subsystem from $k^{\textrm{assoc}}_{x}=2400\ s^{-1}$ to $k^{\textrm{assoc}}_{x}=0.86$ \cite{zabet_2012_subsystem}.}
\label{tab:GRiPTFparams}
\end{table}

\begin{table}
\centering
\begin{tabular}{| l | r | r |r |}
\hline
\textbf{parameter} &  \textbf{$TF_1$} & \textbf{$TF_2$} & \textbf{notation}\\ \hline
association rate to the DNA &  $1.98E8\ s^{-1}$ & $1.98E8\ s^{-1}$ & $k^{\textrm{assoc}}_{x}$\\ \hline
unbinding probability &  $1.0$ & $1.0$ & $P^{\textrm{unbind}}_{x}$\\ \hline
probability to slide left &  $0.0$ & $0.0$ & $P^{\textrm{left}}_{x}$\\ \hline
probability to slide right &  $1.0$ & $1.0$ & $P^{\textrm{right}}_{x}$\\ \hline
probability to dissociate completely when unbinding &  $1.0$ & $1.0$ & $P^{\textrm{jump}}_{x}$\\ \hline
time bound at the target site &  $1336.5\ s$ & $1336.5\ s$ & $\tau^{0}_{x}$\\ \hline
\end{tabular}
\caption{\emph{TF species default parameters for the case when TF molecules perform only 3D diffusion} To match the number of events performed on the DNA as in the case of facilitated diffusion ($1.0E8$ events) we changed the association rate to $1.98E8\ s^{-1}$ and simulated the system for $84600\ s$. The specific waiting time was increased by $4050$ in order to include the average number of 1D events performed during one random walk on the DNA; $\tau_0^*=(s_l^2/2)\tau_0$, where $s_l$ is the sliding length, which we approximated to be $90\ bp$. The rest of the parameters have the values as listed in Table \ref{tab:GRiPTFparams}.}
\label{tab:modelTFparams3D}
\end{table}

\begin{table}
\centering
\begin{tabular}{| l | r | r |r |}
\hline
\textbf{parameter} & \textbf{$TF_1$} & \textbf{$TF_2$} & \textbf{notation}\\ \hline
motif sequence &  \verb+GACTATAGCTTACAAAAAA+ & \verb+CTCTATTATGAGCAACGGT+ & \\ \hline
energetic penalty for mismatch &  $1.122K_BT$ & $1.122K_BT$ & $\varepsilon^{*}_{x}$\\ \hline
nucleotides covered on left &  $1\ bp$ & $1\ bp$ & $TF^{\textrm{left}}_{x}$\\ \hline
nucleotides covered on right &  $1\ bp$ & $1\ bp$ & $TF^{\textrm{right}}_{x}$\\ \hline
\end{tabular}
\caption{\emph{TF species default parameters for the non-uniform landscape}. TFs will cover the same number of nucleotides as in the case of a uniform landscape, namely $21\ bp$. For the non-uniform landscape, we considered a random $20\ Kbp$ DNA sequence using the \emph{E.coli} K-12 nucleotide composition ($A=24.6\%$, $T=24.6\%$, $C=25.4\%$ and  $G=25.4\%$). In addition, the two TF species display four different parameters compared to the uniform landscape case, namely: the motif sequence, the energetic penalty for a mismatch and covered nucleotides to the left and to the right of the motif.  Note that the TF motifs were generated based on the nucleotide compositions of the motifs in \emph{E.coli} ($A=29.2\%$, $T=29.9\%$, $C=20.4\%$ and  $G=20.5\%$). We selected a penalty of $1.122K_BT$ for an energy mismatch, which is within biological plausible parameters ($[1K_BT,3K_BT]$) \cite{gerland_2002} and which ensures that the average binding energy is $-16K_BT$ as in the case of the uniform landscape; see Figure \ref{fig:GRiPnonuniform}\emph{(A-C)}.
The rest of the parameters have the values as listed in Table \ref{tab:GRiPTFparams}.} \label{tab:GRiPnonuniform}
\end{table}

\clearpage

\bibliographystyle{nar}
\bibliography{promoter_arhitecture.bib}

\begin{thebibliography}{10}

\bibitem{hermsen_2006}
Hermsen, R., Tans, S., and ten Wolde, P.~R. (2006)
Transcriptional Regulation by Competing Transcription Factor Modules.
{\em PLoS Computational Biology,} {\bf 2}, 1552--1560.

\bibitem{stormo_2000}
Stormo, G.~D. (2000)
{DNA} binding sites: representation and discovery..
{\em Bioinformatics,} {\bf 16}(1), 16--23.

\bibitem{ackers_1982}
Ackers, G.~K., Johnson, A.~D., and Shea, M.~A. (1982)
Quantitative model for gene regulation by lambda phage repressor.
{\em PNAS,} {\bf 79}, 1129--1133.

\bibitem{djordjevic_2003}
Djordjevic, M., Sengupta, A.~M., and Shraiman, B.~I. (2003)
A Biophysical Approach to Transcription Factor Binding Site Discovery.
{\em Genome Resarch,} {\bf 13}(11), 2381--2390.

\bibitem{bintu_2005_model}
Bintu, L., Buchler, N.~E., Garcia, H.~G., Gerland, U., Hwa, T., Kondev, J., and
  Phillips, R. (2005)
Transcriptional regulation by the numbers: models.
{\em Current Opinion in Genetics and Development,} {\bf 15}, 116--124.

\bibitem{foat_2006}
Foat, B.~C., Morozov, A.~V., and Bussemaker, H.~J. (2006)
Statistical mechanical modeling of genome-wide transcription factor occupancy
  data by {MatrixREDUCE}.
{\em Bioinformatics,} {\bf 22}(14), e141--e149.

\bibitem{roider_2007}
Roider, H.~G., Kanhere, A., Manke, T., and Vingron, M. (2007)
Predicting transcription factor affinities to {DNA} from a biophysical model.
{\em Bioinformatics,} {\bf 23}(2), 134--141.

\bibitem{chu_2009}
Chu, D., Zabet, N.~R., and Mitavskiy, B. (2009)
Models of transcription factor binding: Sensitivity of activation functions to
  model assumptions.
{\em Journal of Theoretical Biology,} {\bf 257}(3), 419--429.

\bibitem{zhao_2009}
Zhao, Y., Granas, D., and Stormo, G.~D. (2009)
Inferring Binding Energies from Selected Binding Sites.
{\em PLoS Computational Biology,} {\bf 5}(12), e1000590.

\bibitem{sadka_2009}
Raveh-Sadka, T., Levo, M., and Segal, E. (2009)
Incorporating nucleosomes into thermodynamic models of transcription
  regulation.
{\em Genome Research,} {\bf 19}, 1480--1496.

\bibitem{wasson_2009}
Wasson, T. and Hartemink, A.~J. (2009)
An ensemble model of competitive multi-factor binding of the genome.
{\em Genome Research,} {\bf 19}, 2101--2112.

\bibitem{hoffman_2010}
Hoffman, M.~M. and Birney, E. (2010)
An effective model for natural selection in promoters.
{\em Genome Resarch,} {\bf 20}(5), 685--692.

\bibitem{kaplan_2011}
Kaplan, T., Li, X.-Y., Sabo, P.~J., Thomas, S., Stamatoyannopoulos, J.~A.,
  Biggin, M.~D., and Eisen, M.~B. (2011)
Quantitative Models of the Mechanisms That Control Genome-Wide Patterns of
  Transcription Factor Binding during Early {Drosophila} Development.
{\em PLoS Genetics,} {\bf 7}(2), e1001290.

\bibitem{simicevic_2013}
Simicevic, J., Schmid, A.~W., Gilardoni, P.~A., Zoller, B., Raghav, S.~K.,
  Krier, I., Gubelmann, C., Lisacek, F., Naef, F., Moniatte, M., and Deplancke,
  B. (2013)
Absolute quantification of transcription factors during cellular
  differentiation using multiplexed targeted proteomics.
{\em Nature Methods,}.

\bibitem{riggs_1970a}
Riggs, A.~D., Bourgeois, S., and Cohn, M. (1970)
The lac represser-operator interaction: {III.} Kinetic studies.
{\em Journal of Molecular Biology,} {\bf 53}(3), 401--417.

\bibitem{kabata_1993}
Kabata, H., Kurosawa, O., I~Arai, M.~W., Margarson, S., Glass, R., and
  Shimamoto, N. (1993)
Visualization of single molecules of RNA polymerase sliding along DNA.
{\em Science,} {\bf 262}(5139), 1561--1563.

\bibitem{blainey_2006}
Blainey, P.~C., van Oijen, A.~M., Banerjee, A., Verdine, G.~L., and Xie, X.~S.
  (2006)
A base-excision DNA-repair protein finds intrahelical lesion bases by fast
  sliding in contact with DNA.
{\em PNAS,} {\bf 103}(15), 5752--5757.

\bibitem{elf_2007}
Elf, J., Li, G.-W., and Xie, X.~S. (2007)
Probing Transcription Factor Dynamics at the Single-Molecule Level in a Living
  Cell.
{\em Science,} {\bf 316}, 1191--1194.

\bibitem{hammar_2012}
Hammar, P., Leroy, P., Mahmutovic, A., Marklund, E.~G., Berg, O.~G., and Elf,
  J. (2012)
The lac Repressor Displays Facilitated Diffusion in Living Cells.
{\em Science,} {\bf 336}(6088), 1595--1598.

\bibitem{berg_1981}
Berg, O.~G., Winter, R.~B., and von Hippel, P.~H. (1981)
Diffusion-driven mechanisms of protein translocation on nucleic acids. 1.
  Models and theory..
{\em Biochemistry,} {\bf 20}(24), 6929--6948.

\bibitem{halford_2004}
Halford, S.~E. and Marko, J.~F. (2004)
How do site-specific {DNA}-binding proteins find their targets?.
{\em Nucleic Acids Research,} {\bf 32}(10), 3040--3052.

\bibitem{mirny_2009}
Mirny, L., Slutsky, M., Wunderlich, Z., Tafvizi, A., Leith, J., and Kosmrlj, A.
  (2009)
How a protein searches for its site on {DNA}: the mechanism of facilitated
  diffusion.
{\em Journal of Physics A: Mathematical and Theoretical,} {\bf 42}, 434013.

\bibitem{zabet_2012_review}
Zabet, N.~R. and Adryan, B. (2012)
Computational models for large-scale simulations of facilitated diffusion..
{\em Molecular BioSystems,} {\bf 8}(11), 2815--2827
{doi:10.1039/C2MB25201E}.

\bibitem{kolomeisky_2011}
Kolomeisky, A.~B. (2011)
Physics of protein-DNA interactions: mechanisms of facilitated target search.
{\em Phys. Chem. Chem. Phys.,} {\bf 13}, 2088--2095.

\bibitem{coppey_2004}
Coppey, M., Benichou, O., Voituriez, R., and Moreau, M. (2004)
Kinetics of Target Site Localization of a Protein on DNA: A Stochastic
  Approach.
{\em Biophysical Journal,} {\bf 87}, 1640--1649.

\bibitem{slutsky_2004b}
Slutsky, M. and Mirny, L.~A. (2004)
Kinetics of Protein-{DNA} Interaction: Facilitated Target Location in
  Sequence-Dependent Potential.
{\em Biophysical Journal,} {\bf 87}, 4021--4035.

\bibitem{sokolov_2005}
Sokolov, I.~M., Metzler, R., Pant, K., and Williams, M.~C. (2005)
{Target Search of N Sliding Proteins on a DNA}.
{\em Biophysical Journal,} {\bf 89}, 895--902.

\bibitem{klenin_2006}
Klenin, K.~V., Merlitz, H., Langowski, J., and Wu, C.-X. (2006)
Facilitated Diffusion of DNA-Binding Proteins.
{\em Phys. Rev. Lett.,} {\bf 96}, 018104.

\bibitem{hu_2006}
Hu, T., Grosberg, A.~Y., and Shklovskii, B.~I. (2006)
How Proteins Search for Their Specific Sites on DNA: The Role of DNA
  Conformation.
{\em Biophysical Journal,} {\bf 90}, 2731--2744.

\bibitem{benichou_2008}
Benichou, O., Loverdo, C., Moreau, M., and Voituriez, R. (2008)
Optimizing intermittent reaction paths.
{\em Phys. Chem. Chem. Phys.,} {\bf 10}(47), 7059--7072.

\bibitem{li_2009}
Li, G.-W., Berg, O.~G., and Elf, J. (2009)
Effects of macromolecular crowding and DNA looping on gene regulation kinetics.
{\em Nature Physics,} {\bf 5}, 294 -- 297.

\bibitem{lomholt_2009}
Lomholt, M.~A., van~den Broek, B., Kalisch, S.-M.~J., Wuite, G. J.~L., and
  Metzler, R. (2009)
Facilitated diffusion with DNA coiling.
{\em PNAS,} {\bf 106}(20), 8204--8208.

\bibitem{loverdo_2009}
Loverdo, C., B\'enichou, O., Voituriez, R., Biebricher, A., Bonnet, I., and
  Desbiolles, P. (2009)
Quantifying Hopping and Jumping in Facilitated Diffusion of DNA-Binding
  Proteins.
{\em Phys. Rev. Lett.,} {\bf 102}, 188101.

\bibitem{meroz_2009}
Meroz, Y., Eliazar, I., and Klafter, J. (2009)
Facilitated diffusion in a crowded environment: from kinetics to stochastics.
{\em Journal of Physics A: Mathematical and Theoretical,} {\bf 42}, 434012.

\bibitem{vukojevic_2010}
Vukojevic, V., Papadopoulos, D.~K., Terenius, L., Gehring, W.~J., and Rigler,
  R. (2010)
Quantitative study of synthetic Hox transcription factor-DNA interactions in
  live cells.
{\em PNAS,} {\bf 107}(9), 4093--4098.

\bibitem{benichou_2011}
Benichou, O., Chevalier, C., Meyer, B., and Voituriez, R. (2011)
Facilitated Diffusion of Proteins on Chromatin.
{\em Phys. Rev. Lett.,} {\bf 106}, 038102.

\bibitem{zhou_2011}
Zhou, H.-X. (2011)
Rapid search for specific sites on {DNA} through conformational switch of
  nonspecifically bound proteins.
{\em PNAS,} {\bf 108}(21), 8651--8656.

\bibitem{zabet_2012_grip}
Zabet, N.~R. and Adryan, B. (2012)
{GRiP}: a computational tool to simulate transcription factor binding in
  prokaryotes.
{\em Bioinformatics,} {\bf 28}(9), 1287--1289.

\bibitem{zabet_2012_model}
Zabet, N.~R. and Adryan, B. (2012)
A comprehensive computational model of facilitated diffusion in prokaryotes.
{\em Bioinformatics,} {\bf 28}(11), 1517--1524.

\bibitem{gama_castro_2011}
Gama-Castro, S., Salgado, H., Peralta-Gil, M., Santos-Zavaleta, A.,
  Muñiz-Rascado, L., Solano-Lira, H., Jimenez-Jacinto, V., Weiss, V.,
  García-Sotelo, J.~S., López-Fuentes, A., Porrón-Sotelo, L.,
  Alquicira-Hernández, S., Medina-Rivera, A., Martínez-Flores, I.,
  Alquicira-Hernández, K., Martínez-Adame, R., Bonavides-Martínez, C.,
  Miranda-Ríos, J., Huerta, A.~M., Mendoza-Vargas, A., Collado-Torres, L.,
  Taboada, B., Vega-Alvarado, L., Olvera, M., Olvera, L., Grande, R., Morett,
  E., and Collado-Vides, J. (2011)
{RegulonDB} version 7.0: transcriptional regulation of \emph{Escherichia coli}
  {K-12} integrated within genetic sensory response units (Gensor Units).
{\em Nucleic Acids Research,} {\bf 39}(suppl 1), D98--D105.

\bibitem{zon_2006}
van Zon, J.~S., Morelli, M.~J., Tanase-Nicola, S., and ten Wolde, P.~R. (2006)
Diffusion of Transcription Factors Can Drastically Enhance the Noise in Gene
  Expression.
{\em Biophysical Journal,} {\bf 91}, 4350--4367.

\bibitem{brackley_2012}
Brackley, C.~A., Cates, M.~E., and Marenduzzo, D. (2012)
Facilitated Diffusion on Mobile {DNA}: Configurational Traps and Sequence
  Heterogeneity.
{\em Phys. Rev. Lett.,} {\bf 109}(16), 168103.

\bibitem{fofanno_2012}
Foffano, G., Marenduzzo, D., and Orlandini, E. (2012)
Facilitated diffusion on confined {DNA}.
{\em Phys. Rev. E,} {\bf 85}(2), 021919.

\bibitem{bauer_2013}
Bauer, M. and Metzler, R. (2013)
In Vivo Facilitated Diffusion Model.
{\em PLoS ONE,} {\bf 8}(1), e53956.

\bibitem{zabet_2013_time}
Zabet, N.~R. and Adryan, B. (2013)
The effects of transcription factor competition on gene regulation.
{\em Frontiers in Genetics,} {\bf 4}(197), 197.

\bibitem{rosenfeld_2005}
Rosenfeld, N., Young, J.~W., Alon, U., Swain, P.~S., and Elowitz, M.~B. (2005)
Gene Regulation at the Single-Cell Level.
{\em Science,} {\bf 307}(5717), 1962--1965.

\bibitem{zabet_2012_subsystem}
Zabet, N.~R. (2012)
System size reduction in stochastic simulations of the facilitated diffusion
  mechanism.
{\em BMC Systems Biology,} {\bf 6}(1), 121.

\bibitem{stewart_1994}
Stewart, W.~J. (1994)
Introduction to the Numerical Solution of Markov Chains,
Princeton University Press, Princeton, New Jersey.

\bibitem{gillespie_1976}
Gillespie, D.~T. (1976)
A general method for numerically simulating the stochastic time evolution of
  coupled chemical reactions.
{\em Journal of Computational Physics,} {\bf 22}(4), 403--434.

\bibitem{gillespie_1977}
Gillespie, D.~T. (1977)
Exact Stochastic Simulation of Coupled Chemical Reactions.
{\em The Journal of Physical Chemistry,} {\bf 81}, 2340--2361.

\bibitem{gillespie_2000}
Gillespie, D.~T. (2000)
The chemical Langevin equation.
{\em Journal of Chemical Physics,} {\bf 113}(1), 297--306.

\bibitem{kolesov_2007}
Kolesov, G., Wunderlich, Z., Laikova, O.~N., Gelfand, M.~S., and Mirny, L.~A.
  (2007)
How gene order is influenced by the biophysics of transcription regulation.
{\em PNAS,} {\bf 104}(35), 13948--13953.

\bibitem{brackley_2013_crowding}
Brackley, C.~A., Cates, M.~E., and Marenduzzo, D. (2013)
Intracellular Facilitated Diffusion: Searchers, Crowders, and Blockers.
{\em Phys. Rev. Lett.,} {\bf 111}(10), 108101.

\bibitem{wunderlich_2008}
Wunderlich, Z. and Mirny, L.~A. (2008)
Spatial effects on the speed and reliability of protein-{DNA} search..
{\em Nucleic Acids Research,} {\bf 36}(11), 3570--3578.

\bibitem{gerland_2002}
Gerland, U., Moroz, J.~D., and Hwa, T. (2002)
Physical constraints and functional characteristics of transcription
  factor-{DNA} interactions.
{\em PNAS,} {\bf 99}(19), 12015--12020.

\bibitem{Fairall1986}
Fairall, L., Rhodes, D., and Klug, A. (Dec, 1986)
Mapping of the sites of protection on a 5 S RNA gene by the Xenopus
  transcription factor IIIA. A model for the interaction..
{\em Journal of Molecular Biology,} {\bf 192}(3), 577--591.

\bibitem{ruusala_1992}
Ruusala, T. and Crothers, D.~M. (1992)
Sliding and intermolecular transfer of the lac repressor: kinetic perturbation
  of a reaction intermediate by a distant DNA sequence.
{\em PNAS,} {\bf 89}(11), 4903--4907.

\bibitem{sharon_2012}
Sharon, E., Kalma, Y., Sharp, A., Raveh-Sadka, T., Levo, M., Zeevi, D., Keren,
  L., Yakhini, Z., Weinberger, A., and Segal, E. (2012)
Inferring gene regulatory logic from high-throughput measurements of thousands
  of systematically designed promoters.
{\em Nature Biotechnology,} {\bf 30}(6), 521--530.

\bibitem{hartigan_1985}
Hartigan, J.~A. and Hartigan, P.~M. (1985)
The Dip Test of Unimodality.
{\em Annals of Statistics,} {\bf 13}(1), 70--84.

\bibitem{diptest_2012}
Maechler, M.
diptest: Hartigan's dip test statistic for unimodality - corrected code (2012)
R package version 0.75-4.

\bibitem{blainey_2009}
Blainey, P.~C., Luo, G., Kou, S.~C., Mangel, W.~F., Verdine, G.~L., Bagchi, B.,
  and Xie, X.~S. (2009)
Nonspecifically bound proteins spin while diffusing along DNA.
{\em Nature Structural \& Molecular Biology,} {\bf 16}(12), 1224 -- 1229.

\bibitem{wang_2012}
Wang, S., Elf, J., Hellander, S., and Lotstedt, P.,
Stochastic reaction-diffusion processes with embedded lower dimensional
  structures.
Technical report,  Department of Information Technology, Uppsala University
  (2012).

\bibitem{He2012}
He, X., Duque, T. S. P.~C., and Sinha, S. (Mar, 2012)
Evolutionary origins of transcription factor binding site clusters..
{\em Mol Biol Evol,} {\bf 29}(3), 1059--1070.

\bibitem{Lusk2010}
Lusk, R.~W. and Eisen, M.~B. (Jan, 2010)
Evolutionary mirages: selection on binding site composition creates the
  illusion of conserved grammars in Drosophila enhancers..
{\em PLoS Genetics,} {\bf 6}(1), e1000829.

\bibitem{Nourmohammad2011}
Nourmohammad, A. and L\"{a}ssig, M. (Oct, 2011)
Formation of regulatory modules by local sequence duplication..
{\em PLoS Computational Biology,} {\bf 7}(10), e1002167.

\bibitem{Lu2007}
Lu, P., Vogel, C., Wang, R., Yao, X., and Marcotte, E.~M. (Jan, 2007)
Absolute protein expression profiling estimates the relative contributions of
  transcriptional and translational regulation..
{\em Nature Biotechnology,} {\bf 25}(1), 117--124.

\bibitem{cox_2007}
Cox~III, R.~S., Surette, M.~G., and Elowitz, M.~B. (2007)
Programming gene expression with combinatorial promoters.
{\em Molecular Systems Biology,} {\bf 3}, 145.

\bibitem{Dadiani2013}
Dadiani, M., {van Dijk}, D., Segal, B., Field, Y., Ben-Artzi, G., Raveh-Sadka,
  T., Levo, M., Kaplow, I., Weinberger, A., and Segal, E. (May, 2013)
Two DNA-encoded strategies for increasing expression with opposing effects on
  promoter dynamics and transcriptional noise..
{\em Genome Research,}.

\bibitem{khoueiry_2010}
Khoueiry, P., Rothbächer, U., Ohtsuka, Y., Daian, F., Frangulian, E., Roure,
  A., Dubchak, I., and Lemaire, P. (2010)
A cis-Regulatory Signature in Ascidians and Flies, Independent of Transcription
  Factor Binding Sites.
{\em Current Biology,} {\bf 20}(9), 792 -- 802.

\bibitem{janga_2008}
Janga, S.~C., Collado-Vides, J., and Babu, M.~M. (2008)
Transcriptional regulation constrains the organization of genes on eukaryotic
  chromosomes.
{\em PNAS,} {\bf 105}(41), 15761--15766.

\bibitem{shen_orr_2002}
Shen-Orr, S.~S., Milo, R., Mangan, S., and Alon, U. (2002)
Network motifs in the transcriptional regulation network of \emph{Escherichia
  coli}.
{\em Nature Genetics,} {\bf 31}, 64--68.

\bibitem{zabet_2009}
Zabet, N.~R. and Chu, D.~F. (2010)
Computational limits to binary genes.
{\em Journal of the Royal Society Interface,} {\bf 7}, 945--954.

\bibitem{he_2009}
He, X., Chen, C.-C., Hong, F., Fang, F., Sinha, S., Ng, H.-H., and Zhong, S.
  (2009)
A Biophysical Model for Analysis of Transcription Factor Interaction and
  Binding Site Arrangement from Genome-Wide Binding Data.
{\em PLoS ONE,} {\bf 4}(12), e8155.

\bibitem{cheng_2013}
Cheng, Q., Kazemian, M., Pham, H., Blatti, C., Celniker, S.~E., Wolfe, S.~A.,
  Brodsky, M.~H., and Sinha, S. (2013)
Computational Identification of Diverse Mechanisms Underlying Transcription
  Factor-{DNA} Occupancy.
{\em PLoS Genetics,} {\bf 9}(8), e1003571.

\bibitem{weindl_2007}
Weindl, J., Hanus, P., Dawy, Z., Zech, J., Hagenauer, J., and Mueller, J.~C.
  (2007)
Modeling {DNA}-binding of {Escherichia coli} $ \sigma^{70}$ exhibits a
  characteristic energy landscape around strong promoters.
{\em Nucleic Acids Research,} {\bf 35}(20), 7003--7010.

\bibitem{weindl_2009}
Weindl, J., Dawy, Z., Hanus, P., Zech, J., and Mueller, J.~C. (2009)
Modeling promoter search by E. coli RNA polymerase: One-dimensional diffusion
  in a sequence-dependent energy landscape.
{\em Journal of Theoretical Biology,} {\bf 259}(3), 628--634.

\bibitem{riley_2006}
Riley, M., Abe, T., Arnaud, M.~B., Berlyn, M.~K., Blattner, F.~R., Chaudhuri,
  R.~R., Glasner, J.~D., Horiuchi, T., Keseler, I.~M., Kosuge, T., Mori, H.,
  Perna, N.~T., Plunkett, G., Rudd, K.~E., Serres, M.~H., Thomas, G.~H.,
  Thomson, N.~R., Wishart, D., and Wanner, B.~L. (2006)
Escherichia coli K-12: a cooperatively developed annotation snapshot - 2005.
{\em Nucleic Acids Research,} {\bf 34}(1), 1--9.

\bibitem{berg_1987}
Berg, O.~G. and von Hippel, P.~H. (1987)
Selection of {DNA} Binding Sites by Regulatory Proteins Statistical-mechanical
  Theory and Application to Operators and Promoters.
{\em Journal of Molecular Biology,} {\bf 193}(4), 723--750.

\bibitem{stormo_1998}
Stormo, G.~D. and Fields, D.~S. (1998)
Specificity, free energy and information content in protein-{DNA} interactions.
{\em Trends in Biochemical Sciences,} {\bf 23}(3), 109--113.

\end{thebibliography}

\end{document}